\documentclass[aps,prb,twocolumn,showpacs,superscriptaddress,reprint,longbibliography]{revtex4-1}

\usepackage{amsmath, amsfonts, amssymb}
\usepackage{enumerate}
\usepackage{graphicx}
\usepackage{color}
\usepackage[breaklinks]{hyperref}
\usepackage[hyphenbreaks]{breakurl}
\usepackage{paralist}
\usepackage{multirow}

\newcommand{\eref}[1]{Eq.~(\ref{#1})}
\newcommand{\fref}[1]{Fig.~\ref{#1}}

\newcommand{\sref}[1]{Sec.~\ref{#1}}

\newcommand{\ti}{\hat{\tau}_{0}}
\newcommand{\tx}{\hat{\tau}_{1}}
\newcommand{\ty}{\hat{\tau}_{2}}
\newcommand{\tz}{\hat{\tau}_{3}}

\newcommand{\mean}[1]{\langle #1 \rangle}

\DeclareMathOperator{\sgn}{sgn}

\begin{document}

\title{Current fluctuations in unconventional superconductor junctions with impurity scattering}

\newcommand{\wuerzburg}{Institute for Theoretical Physics and Astrophysics,
	University of W\"{u}rzburg, D-97074 W\"{u}rzburg, Germany}
\newcommand{\aalto}{Department of Applied Physics,
	Aalto University, FIN-00076 Aalto, Finland}
\newcommand{\nagoya}{Department of Applied Physics, Nagoya University, Nagoya 464-8603, Japan}
\newcommand{\sapporo}{Department of Applied Physics, Hokkaido University, Sapporo 060-8628, Japan}
\newcommand{\manchester}{National Graphene Institute, University of Manchester, Booth St E, M13 9PL, Manchester, UK}

\author{Pablo Burset}
\affiliation{\wuerzburg}
\affiliation{\nagoya}
\affiliation{\aalto}

\author{Bo Lu}
\affiliation{\nagoya}
\affiliation{\manchester}

\author{Shun Tamura}
\affiliation{\nagoya}
 
\author{Yukio Tanaka}
\affiliation{\nagoya}
 
\date{\today}

\pacs{73.23.-b,74.20.Rp,74.45.+c,74.50.+r}


\begin{abstract}
The order parameter of bulk two-dimensional superconductors is classified as \textit{nodal}, if it vanishes for a direction in momentum space, or \textit{gapful} if it does not. Each class can be topologically nontrivial if Andreev bound states are formed at the edges of the superconductor. 
Non-magnetic impurities in the superconductor affect the formation of Andreev bound states and can drastically change the tunneling spectra for small voltages. 
Here, we investigate the mean current and its fluctuations for two-dimensional tunnel junctions between a normal-metal and unconventional superconductors by solving the quasi-classical Eilenberger equation self-consistently, including the presence of non-magnetic impurities in the superconductor. 
As the impurity strength increases, we find that superconductivity is suppressed for almost all order parameters since 
\begin{inparaenum}[i)]
	\item at zero applied bias, the effective transferred charge calculated from the noise-current ratio tends to the electron charge $e$ and 
	\item for finite bias, the current-voltage characteristics follows that of a normal state junction. 
\end{inparaenum}
There are notable exceptions to this trend. 
First, gapful nontrivial (chiral) superconductors are very robust against impurity scattering due to the linear dispersion relation of their surface Andreev bound states. 
Second, for nodal nontrivial superconductors, only $p_x$-wave pairing is almost immune to the presence of impurities due to the emergence of odd-frequency $s$-wave Cooper pairs near the interface. 
Owing to their anisotropic dependence on the wave vector, impurity scattering is an effective pair breaking mechanism for the rest of nodal superconductors. 
All these behaviors are neatly captured by the noise-current ratio, providing a useful guide to find experimental signatures for unconventional superconductivity. 
\end{abstract}

\maketitle

\section{Introduction}
The symmetry of the superconducting order parameter is crucial to determine many properties of a superconductor. The majority of superconductors feature a conventional spin-singlet $s$-wave pair potential. 
Any deviation from this pair potential, be it spin-triplet states or higher harmonics like $p$-wave or $d$-wave, is considered unconventional\cite{Sigrist_RMP}. 
One of the most interesting consequences of unconventional pairings is the formation of surface Andreev bound states (SABS) when the pair potential changes sign on the Fermi surface\cite{Zwicknagl_1981,Bruder_1990,Hu_1994,Tanaka_1995,Kashiwaya_2000}. The formation of SABS is related with the emergence of a zero bias peak (ZBP) in the tunnel conductance\cite{Tanaka_1995,Kashiwaya_2000}. 
While conventional $s$-wave pairing is robust against non-magnetic impurities\cite{Anderson_1959}, many unconventional pairings are fragile owing to their anisotropic dependence on the wave vector\cite{Balatsky_RMP}. 
Some SABS have a topological origin and would be protected against imperfections or impurities\cite{Schnyder_2008,Sato_2009,Sato_2010,Schnyder_2010,Qi_RMP,Mizushima_2015}. However, impurity scattering reduces or completely suppresses the ZBP for many cases, making it difficult to detect unconventional pairing symmetries from conductance measurements only\cite{Lu_2016}. 
To go beyond dc conductance, it is interesting to study the non-equilibrium current fluctuations or shot noise\cite{Blanter_2000}. 
The shot noise reveals the effective charge transferred in a given tunneling process through the noise-current ratio. 
For example, the effective charge of a tunnel junction between a normal metal and a superconductor is doubled, revealing the uncorrelated transfer of Cooper pairs due to Andreev processes\cite{Khlus_1987,*noiseNS_1994,*Beenakker_1994,Anantram-Datta,Klapwijk_1997}. 

In this work we study the current, shot noise and noise-current ratio of normal-metal--superconductor junctions, including the effect of non-magnetic impurity scattering in the superconductor, for the most representative two-dimensional unconventional order parameters. 
Depending on the shape of the order parameter in reciprocal space, superconductors in two dimensions can be classified into two groups: 
\begin{inparaenum}[(i)] 
	\item Gapful superconductors with a finite order parameter and 
	\item Nodal superconductors where the order parameter vanishes in a given direction. 
\end{inparaenum}
At the same time, each order parameter can be topologically nontrivial or trivial depending on whether SABS appear or not. 
For example, the conventional spin-singlet $s$-wave state belongs to the \textit{gapful trivial} group. 
Chiral superconductors\cite{Chiral_SC_2016} are also gapped in the bulk, but feature SABS with a linear dispersion relation; they thus belong to the \textit{gapful nontrivial} group. 
Sr$_2$RuO$_4$ is a strong candidate for chiral spin-triplet $p$-wave superconductor\cite{Mackenzie_2003,*Maeno_2012,*Kallin_2012}. 
Experiments have failed to detect the predicted spontaneous edge current in Sr$_2$RuO$_4$, suggesting that the chiral symmetry could be on a higher harmonic, like $d$- or $f$-wave\cite{Kallin_2014,*Masaki_2015,*Simon_2015}. Chiral pairing states have also been proposed for other systems, including graphene\cite{Chiral_graphene_2014}. 
Chiral superconductors are currently attracting a lot of attention since their topologically nontrivial edge states, which display a linear dependence on the momentum, are a condensed matter realization of Majorana states\cite{Read_2000,Fujimoto_2008,Sau_2010,*Lutchyn_2010}. 
On the other hand, nodal superconductors with a vanishing order parameter feature SABS with a flat dispersion relation at their edges\cite{Yokoyama_2011}. 
Nodal superconductivity naturally appears in high-Tc cuprates\cite{Tanaka_1995,*Kashiwaya_1995,Kashiwaya_2000} ($d$-wave) and noncentrosymmetric superconductors\cite{Yada_2010,*Yada_2011}. It can also be engineered by proximity effect from a conventional superconductor in materials with strong spin-orbit coupling\cite{Vedral_2013,Ikegaya_2015} ($p$-wave). 
Assuming that the junction lies along the $x$-direction, \textit{nodal trivial} groups include $p_y$- and $d_{x^2-y^2}$-wave, while \textit{nodal nontrivial} correspond to $p_x$, $d_{xy}$ and similar\footnotemark[1]. 

This paper is organized as follows. We describe our model and present the main definitions for the transport observables in \sref{sec:model}. In \sref{sec:ball}, we present an exhaustive collection of transport results for ballistic junctions with unconventional superconductors. Here, our model reproduces many well-known results from previous works and we discuss the most representative behavior of the different pairing symmetries. Next, in \sref{sec:imp} we show the main results of this work as we discuss the effect of impurities on the current, shot noise and noise-current ratio of unconventional superconductors. We report our conclusions in \sref{sec:conc}. 

\footnotetext[1]{We are considering the situation where the nodal direction lies along the $x$-direction. In a more general case, there can be an angle between the nodal direction and the $x$-axis. A slightly tilted $p_y$-wave or $d_{x^2-y^2}$-wave pairing is then nontrivial, but it will behave in a similar way as the trivial cases studied here. We are only interested in the representative behavior for two-dimensional superconductors, so we will not consider such cases here. }


\section{Model \label{sec:model}} 
We consider a two-dimensional normal-metal--superconductor junction where transport takes place along the $x$ direction and set the interface at $x\!=\!0$. We thus parametrize the conserved transverse component of the wave vector $k_y$ using the angle of incidence $\theta\!=\!\sin^{-1}(k_y/k_F)$, with $k_F$ the Fermi wave vector. Depending on the direction of propagation of the quasiparticles, we define the angles $\theta_{+}\!\equiv\!\theta\!\in\![-\pi/2,\pi/2]$ and $\theta_{-}\!=\!\pi-\theta$. 
We model the scattering at the interface using a $\delta$-function potential $V(x)\!=\!Z(\hbar^2 k_F/2m)\delta(x)$, with $Z$ the dimensionless barrier strength and $m$ the electron mass. 
We assume a clean normal metal ($x\!<\!0$) and a uniform distribution of non-magnetic impurities in the superconducting region ($x\!>\!0$) with induced self-energy $\hat{a}(x)$. The superconducting order parameter is given by $\hat{\Delta}(\theta_{\alpha},x)$, with $\alpha\!=\!\pm$. In the normal region, we take $\hat{a}(x\!<\!0)\!=\!\hat{\Delta}(\theta_{\alpha},x\!<\!0)\!=\!0$. 
For a spin-degenerate system, the quasi-classical Green's function\cite{Rainer_1983,Rammer_RMP,Sauls_1988,Nagai_1989} $\hat{g}^{\alpha\alpha}(i\omega_n,\theta_\alpha,x)$ for Matsubara frequency $\omega_n\!=\!(2n+1)\pi T$, where $T$ is the temperature and $n$ an integer, is a $2\!\times\!2$ matrix in particle-hole space that satisfies the Eilenberger equation\cite{Eilenberger_1968}
\begin{equation}\label{eq:eilenberger}
iv_{Fx}\partial_x \hat{g}^{\alpha\alpha} + \alpha[i\omega_n\tz-\hat{\Delta}(\theta_{\alpha},x)\tz-\hat{a}(x),\hat{g}^{\alpha\alpha}]=0 .
\end{equation}
Here, $v_{Fx}\!=\!v_F\cos\theta$ is the $x$ component of the Fermi velocity $v_F$ and the particle-hole space is spanned by Pauli matrices $\hat{\tau}_{0,1,2,3}$, with $\ti$ the identity matrix. The quasi-classical Green's function is normalized as $(\hat{g}^{\alpha\alpha})^2\!=\!-1$. 

\begin{table*}
	\begin{tabular}{ c c c c c c  c c  c c  c c  c c }
		\hline\hline
		\multirow{2}{*}{type} & \multirow{2}{*}{wave} & \multirow{2}{*}{$\chi_R(\theta_{\pm})$} & \multirow{2}{*}{$\chi_I(\theta_{\pm})$} & \multirow{2}{*}{node}  & \multirow{2}{*}{SABS} & \multicolumn{2}{c}{$\sigma_S/\sigma_N$} &
		\multicolumn{2}{c}{$P_S/S_N$} & \multicolumn{2}{c}{$P_S/(2e\sigma_S)$} & \multicolumn{2}{c}{$I_{\mathrm{exc}}$} 
		\\ 
		& & & & & & bal & impurity & bal. & impurity & bal. & impurity & bal. & impurity
		\\	 
		\hline\noalign{\smallskip}
		\parbox[c]{1.3cm}{1.$\!$ Gapful trivial}  & $s$ & $1$ & $0$ & $\times$ & $\times$ & $\rightarrow0$ & $\rightarrow0$ &  $\rightarrow0$ & $\rightarrow0$ & $2$ & $\rightarrow1$ & $0$ & $0$
		\\ 
		\multirow{3}{*}{\parbox[c]{1.3cm}{2.$\!$ Gapful non-trivial}} & chiral-$p$ & $\pm\cos\theta$ & $\sin\theta$ & $\times$ & linear & $\sim 1$ & $\sim1$ &  $\sim 1$ & $\sim 1$ & $\lesssim1$ & $\sim1$ & $0$ & $0$ 
		\\
		& chiral-$d$ & $\cos2\theta$ & $\pm\sin2\theta$ & $\times$ & linear & $\sim 1/2$ & $\gtrsim1/2$ &  $\sim 1/2$ & $\sim 1/2$ & $\gtrsim1$ & $\sim1$ & $0$ & $0$
		\\
		& chiral-$f$ & $\pm\cos3\theta$ & $\sin3\theta$ & $\times$ & linear & $\sim 1/2$ & $\gtrsim1/2$ &  $\sim 1/2$ & $\sim 1/2$ & $\gtrsim1$ & $\sim1$ & $0$ & $0$
		\\ 
		\multirow{4}{*}{\parbox[c]{1.3cm}{3.$\!$ Nodal trivial}} & \multirow{2}{*}{$p_y$} & \multirow{2}{*}{$\sin\theta$} & \multirow{2}{*}{$0$} & \multirow{2}{*}{$\checkmark$} & \multirow{2}{*}{$\times$} & \multirow{2}{*}{$>0$} & $\sim1/10$ (B) &  \multirow{2}{*}{$\ll 1$} & $\ll 1$ (B) & \multirow{2}{*}{$\sim2$} & $\gtrsim3/2$ (B) & \multirow{2}{*}{$0$} & \multirow{2}{*}{$0$} 
		\\
		&  &  &  &  &  &  & $\sim1/2$ (U) &  & $\sim 1/2$ (U) &  & $\sim1$ (U) & & 
		\\
		& \multirow{2}{*}{$d_{x^2-y^2}$} & \multirow{2}{*}{$\cos2\theta$} & \multirow{2}{*}{$0$} & \multirow{2}{*}{$\checkmark$} & \multirow{2}{*}{$\times$} & \multirow{2}{*}{$>0$} & $\gtrsim1/10$ (B) & \multirow{2}{*}{$\ll 1$} & $\ll 1$ (B) & \multirow{2}{*}{$\lesssim2$} & $\lesssim2$ (B) & \multirow{2}{*}{$0$} & \multirow{2}{*}{$0$} 
		\\
		&  &  &  &  &  &  & $\sim1/2$ (U)  &  & $\sim 1/2$ (U) &  & $\sim1$ (U) & & 
		\\	
		\multirow{3}{*}{\parbox[c]{1.3cm}{4.$\!$ Nodal non-trivial}} & $p_x$ & $\pm\cos\theta$ & $0$ & $\checkmark$ & flat & ZBP & ZBP &  $0$ & $0$ & $0$ & $0$ & $I_{\mathrm{exc}}^{\mathrm{ball}}$ & $I_{\mathrm{exc}}^{\mathrm{ball}}$
		\\
		& $d_{xy}$ & $\pm\sin2\theta$ & $0$ & $\checkmark$ & flat & ZBP & $\sim 2$ &  $0$ & $\sim3/2$ & $0$ & $\lesssim1$ & $I_{\mathrm{exc}}^{\mathrm{ball}}$ & $\rightarrow0$
		\\
		& $f_x$ & $\pm\cos3\theta$ & $0$ & $\checkmark$ & flat & ZBP & $\sim2$ &  $0$ & $\sim3/2$ & $0$ & $\lesssim1$ & $I_{\mathrm{exc}}^{\mathrm{ball}}$ & $\rightarrow0$
		\\ 
		\hline
	\end{tabular}
	\caption{Symmetry of the superconducting pairing, transport results for $E\!=\!0$ and $Z\!=\!5$, and excess current. The symbol $\checkmark$ ($\times$) represents ``presence of \dots" (``absence of \dots"). Zero-bias peak (ZBP) indicates the case where $\sigma_S\!\gg\!\sigma_N$. Results in the ballistic (bal.) and impurity regimes are taken with $1/(2\tau\Delta_b)\!=\!0$ and $0.2$, respectively. Born (B) and Unitary (U) limits are calculated with $\sigma\!=\!0$ and $\sigma\!=\!0.99$, respectively. }
	\label{table}
\end{table*}

To account for unconventional superconductivity in the rightmost region ($x\!>\!0$), we use the notation
\begin{equation}\label{eq:order-param}
\hat{\Delta}(\theta_{\alpha},x)\equiv \left[ \Delta_R(x)\chi_R(\theta_{\alpha})\tx-\Delta_I(x)\chi_I(\theta_{\alpha})\ty \right]\Theta(x),
\end{equation}
with $\Theta(x)$ the Heaviside function. The subindices $R,I$ refer to the real or imaginary part of the order parameter. 
We choose the global $U(1)$ gauge so that the order parameter is real for non-chiral superconductors or it is proportional to a cosine function of the angle for chiral ones. The resulting form factors $\chi_{R,I}(\theta_{\alpha})$ are enumerated in Table~\ref{table}. 

The spatial dependence of the order parameter is determined self-consistently in terms of the quasi-classical Green's function, namely\cite{Lu_2016,Tanaka_2007c}
\begin{align}
\Delta_R(x){}=& \frac{ 2T\sum\limits_{n\geq 0} \mean{ \mean{ \chi_R(\theta_{\alpha}) \left[ \hat{g}^{\alpha\alpha}(\omega_n,\theta_\alpha,x)\right]_{12} } }_{\theta} }{ \ln\frac{T}{T_c} + \sum\limits_{n\geq 0} \frac{1}{n-1/2} } , \\ 
\Delta_I(x){}=& -i\frac{ 2T\sum\limits_{n\geq 0} \mean{\mean{ \chi_I(\theta_{\alpha}) \left[ \hat{g}^{\alpha\alpha}(\omega_n,\theta_\alpha,x)\right]_{12} } }_{\theta} }{ \ln\frac{T}{T_c} + \sum\limits_{n\geq 0} \frac{1}{n-1/2} } ,
\end{align}
with the angle average defined as $\mean{\mean{f(\theta_\alpha)}}_{\theta} \!=\!\sum_{\alpha} \int_{-\pi/2}^{\pi/2}\!\mathrm{d}\theta f(\theta_\alpha)$. The sums include a cutoff $n_{max}$, defined as the maximum integer that satisfies $n_{max}\!\leq\! \omega_D/(2\pi T)$. $T_c$ is the critical temperature of the bulk superconductor and $\omega_D\!=\!2\pi T_c$ is the Debye frequency, ignoring thermodynamic phenomena. In the bulk of the superconductor, i.e., deep inside the superconducting region, a finite order parameter fulfills $\Delta_{R,I}(x\rightarrow\infty)\!\equiv\!\Delta_b$. 

Following Ref.~\onlinecite{Lu_2016}, the self-energy for the distribution of non-magnetic impurities is written as $\hat{a}=\sum_{j=1}^{3}a_{j}\hat{\tau}_{j}$, with 
\begin{equation}\label{eq:self-energy}
a_j(\omega_n,x)=\frac{ \frac{-1}{2\tau}\frac{1}{1-\sigma} \mean{\mean{ g_{j}^{\alpha\alpha}(\omega_n,x) }}_{\theta} }{1-\frac{1}{1-\sigma}\sum_{j}\left[\mean{\mean{ g_{j}^{\alpha\alpha}(\omega_n,x) }}_{\theta}\right]^2 } ,
\end{equation}
where $1/\tau$ and $\sigma$ are the normal scattering rate and the strength of a single impurity potential, respectively. 

For the numerical calculations, it is useful to express the Green's function in terms of the Riccati parameters\cite{Nagato_1993,Kazumi_1995,Ozana_2000} as
\begin{equation}
\hat{g}^{\alpha\alpha}= \frac{i\alpha}{1-\mathcal{G}^S_{\alpha}\mathcal{F}^S_{\alpha}} \begin{bmatrix} 1+\mathcal{G}^S_{\alpha}\mathcal{F}^S_{\alpha} & 2i\mathcal{F}^S_{\alpha} \\ 2i\mathcal{G}^S_{\alpha} & -1-\mathcal{G}^S_{\alpha}\mathcal{F}^S_{\alpha} \end{bmatrix} , 
\end{equation}
where $\mathcal{G}^S_{\alpha}(\omega_n,\theta_{\alpha},x)$ and $\mathcal{F}^S_{\alpha}(\omega_n,\theta_{\alpha},x)$ satisfy the equations
\begin{align}
\alpha v_{Fx}\partial_x \mathcal{G}^S_{\alpha} {}=& 2(\omega_n-ia_3)\mathcal{G}^S_{\alpha} +\Lambda_{1}^{\alpha} \left(\mathcal{G}^S_{\alpha}\right)^2 -\Lambda_{2}^{\alpha}, \\
\alpha v_{Fx}\partial_x \mathcal{F}^S_{\alpha} {}=& -2(\omega_n-ia_3)\mathcal{F}^S_{\alpha} +\Lambda_{2}^{\alpha} \left(\mathcal{F}^S_{\alpha}\right)^2 -\Lambda_{1}^{\alpha} , 
\end{align}
with 
\begin{equation}
\Lambda_{1,2}^{\alpha}=\Delta_R\chi_R(\theta_{\alpha}) +ia_2 \pm i \left[ \Delta_I\chi_I(\theta_{\alpha}) + ia_1 \right] .
\end{equation}
Finally, at the interface, we set the boundary conditions\cite{Nagai_1995,Shiba_1996,Eschrig_2000,Fogelstrom_2000,Sauls_2004}
\begin{equation}
\mathcal{F}_{\alpha}^{S}(x=0)\rightarrow
\frac{(1-\sigma_{n\theta})^{\alpha\sgn(\omega_n)}}{\mathcal{G}_{-\alpha}^{S}(x=0)},
\end{equation}
with $\sigma_{n\theta}\!=\!4\cos^{2}\theta/\left( Z^{2}\!+\!4\cos^2\theta \right)$ the normal state angle-dependent transmission. 

Following the scattering formalism\cite{BTK}, the Andreev ($a$) and normal ($b$) reflection amplitudes at the interface are given by 
\begin{align}
a{}=&\frac{i\mathcal{\bar{G}}_{+}^{S}}{ 1 + \sigma_{n\theta} \left(1 - \mathcal{\bar{G}}_{+}^{S}/\mathcal{\bar{G}}_{-}^{S}\right) }  , \\
b{}=&\frac{Z}{2i\cos\theta-Z}\frac{\frac{1}{\sigma_{n\theta}} \left(1 - \mathcal{\bar{G}}_{+}^{S}/\mathcal{\bar{G}}_{-}^{S}\right)}{{ 1 + \sigma_{n\theta}\left(1 - \mathcal{\bar{G}}_{+}^{S}/\mathcal{\bar{G}}_{-}^{S}\right) }}  ,
\end{align}
with $\mathcal{\bar{G}}_{\alpha}^{S}\!=\!\mathcal{G}_{\alpha}^{S}\left(
E,\theta_{\alpha},x\!=\!0\right)$ and $E\!>\!0$ the real excitation energy of an incident quasiparticle. 

Using the reflection amplitudes, we define the differential conductance\cite{Tanaka_1995,Kashiwaya_2000}  
\begin{equation}
\sigma_{S}(E)= \frac{2e^2}{h} \int_{-\pi/2}^{\pi/2}\! \mathrm{d}\theta\cos\theta \left( 1 - \left\vert b\right\vert^2 + \left\vert
a\right\vert^2 \right) ,
\label{eq:cond}
\end{equation}
and differential noise power\cite{Anantram-Datta} 
\begin{align}
P_{S}(E){}=& \frac{4e^3}{h} \int_{-\pi/2}^{\pi/2}\! \mathrm{d}\theta\cos\theta \left[ \left\vert a\right\vert^2\left(1-\left\vert a\right\vert^2\right)  \right. \nonumber \\ & \left. + \left\vert b\right\vert^2\left(1-\left\vert b\right\vert^2\right) +2 \left\vert a\right\vert^2 \left\vert b\right\vert^2 \right]  . 
\label{eq:npow}
\end{align}
In the normal state, the differential conductance and noise power are respectively defined as
\begin{align*}
\sigma_{N}=R_N^{-1}={}& \frac{e^2}{h}\int_{-\pi/2}^{\pi/2}\!\mathrm{d}\theta \cos\theta\sigma_{n\theta} , \\
S_{N}={}& \frac{2e^3}{h} \int_{-\pi/2}^{\pi/2}\!\mathrm{d}\theta \cos\theta\sigma_{n\theta}(1\!-\!\sigma_{n\theta}) .
\end{align*}  
The zero-temperature current and shot noise are then obtained integrating Eqs. (\ref{eq:cond}) and (\ref{eq:npow}) for a finite voltage, respectively, 
\begin{align}
I_S(eV) {}=& \int_{0}^{eV}\sigma_{S}(E) \mathrm{d}E  ,
\label{eq:curr} \\ 
S_S(eV) {}=& \int_{0}^{eV} P_{S}(E) \mathrm{d}E  ,
\label{eq:snos}
\end{align}
with $V$ the voltage drop at the NS interface\footnotemark[2]. 

\footnotetext[2]{In the self-consistent evaluation of the pairing states, we have not included thermodynamic phenomena. Accordingly, in the numerical calculations, we choose a sufficiently small temperature $T=0.05T_c$. }

\begin{figure*}
	\includegraphics[width=1.\textwidth]{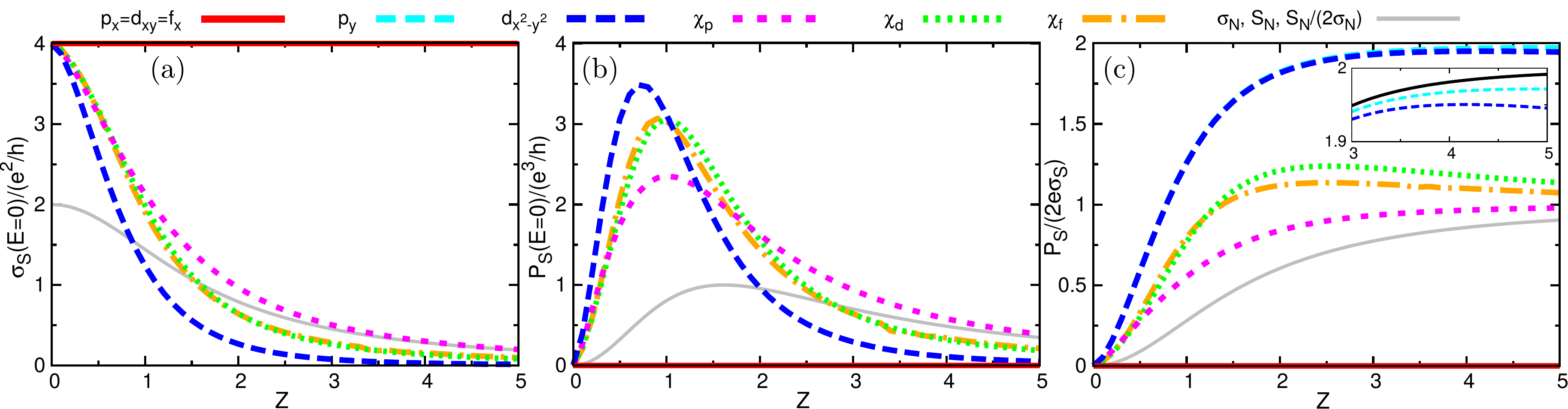}
	\caption{\label{fig:ball_Z} 
		Transport results without impurities. 
		Zero-bias differential conductance (a), noise power (b), and their ratio (c) as a function of the barrier strength $Z$ for different unconventional pairing symmetries. The gray lines show the corresponding results for a junction in the normal state. The inset in (c) compares $p_y$- and $d_{x^2-y^2}$-wave symmetries with $s$-wave case (black solid line). }
\end{figure*}


\section{Ballistic junction \label{sec:ball}} 
In this section, we use our model for the study of ballistic (impurity-free) normal-metal--superconductor junctions with a barrier controlling the interface transmission. 
The following results for conductance, shot noise and noise-current ratio are gathered in Table~\ref{table} under the columns ``ballistic''. 

In the limit of transparent junction, with $Z\!=\!0$, all types of superconductor feature a perfect Andreev reflection at the interface. 
Consequently, the differential conductance is a constant with twice the value of the normal state conductance for small applied bias voltage compared to the bulk gap [see \fref{fig:ball_Z}(a)]. To clearly distinguish between the different types of superconductor, one must make use of the tunnel conductance, opening the possibility of normal backscattering at the interface. 
As we approach the tunnel limit, $Z\!\gg\!1$, the zero-energy conductance for each type of superconductor becomes different featuring three illustrative behaviors. 

Nodal nontrivial superconductors, with $p_x$-, $d_{xy}$-, or $f_x$-wave symmetry, feature a perfect Andreev reflection independently of the barrier strength. Since the normal state conductance is reduced by increasing $Z$, the normalized tunnel conductance $\sigma_S/\sigma_N$ prominently displays a zero-bias peak. 

For gapful nontrivial (chiral) superconductors, the conductance reduces to a finite value, slightly over $\sigma_N$ for the chiral $p$-wave case or comparable to $\sigma_N/2$ for the rest of chiral pairing states. 

For trivial superconductors, both nodal and gapful, the conductance is reduced well below the normal state conductance $\sigma_N$. The resulting normalized tunnel conductance is strongly suppressed for small energies, featuring a (U-) V-shape profile for (gapful) nodal pairing states. 
In the gapful trivial case the conductance tends to zero, while for the nodal trivial cases it tends to a finite but small  value\cite{Lu_2016}. 

The tunnel conductance in the ballistic limit is thus a useful tool to explore the symmetry of the superconducting pairing. However, the height of the zero-bias peak and the gap suppression are very sensitive to the barrier strength and are also rounded by temperature effects\cite{Lu_2016}. Therefore, tunnel conductance experiments can sometimes be ambiguous. 
Charge fluctuations of the current provide an extra layer of information on the symmetry of the pairing potential. 
For a ballistic junction, the noise power at zero temperature can also be interpreted in terms of the reflection processes only\cite{Khlus_1987,*noiseNS_1994,*Beenakker_1994,Anantram-Datta}. For energies below the gap, the integrand of \eref{eq:npow} reduces to $4\left\vert a\right\vert^2 (1-\left\vert a\right\vert^2)$. Consequently, for perfect Andreev reflection ($\left\vert a\right\vert^2\!=\!1$) or in the absence of it ($\left\vert a\right\vert^2\!=\!0$), the noise power is zero. Therefore, nodal nontrivial superconductors are always noiseless at zero energy independently of the barrier strength\cite{Zhu-Ting_1999,Inoue_2000}, as shown in \fref{fig:ball_Z}(b). 
The noise power for the rest of pairing states develops a maximum between the transparent and tunnel limits. 
The maxima for each pairing occur for different values of the barrier strength. However, it would be pointless to use this to experimentally identify each symmetry since the barrier $Z$ is a fitting parameter that accounts for many possible sources of interfacial scattering\cite{BTK}. 

A more clear distinction between all superconductors is given by the noise-current ratio, shown in \fref{fig:ball_Z}(c), which determines the effective charge transferred at the interface.
Since nodal nontrivial superconductors are noiseless independently of the barrier strength, their ratio is also zero. 
Gapped nontrivial (chiral) superconductors have an effective charge equal to the electron charge in the tunnel limit\cite{Inoue_2000}. In this case, the conducting channels are a superposition of modes with a strong Andreev reflection amplitude (i.e., those with angle of incidence $|\theta|\!\gtrsim\!0$ that feature a linear SABS) and others with strong normal backscattering (for $|\theta|\!\lesssim\!\pi/2$). 
For trivial superconductors, gapful or nodal, the effective transferred charge approaches $2e$ in the tunnel regime, indicating the transfer of a Cooper pair at the junction. Small differences between $s$-, $p_y$- and $d_{x^2-y^2}$-wave superconductors appear in the tunnel limit, as shown in the inset of \fref{fig:ball_Z}(c). However, they do not affect the general behavior of the ratio. 

In summary, there are three general trends for ballistic junctions that are neatly captured in the noise-current ratio in the tunnel limit: \begin{inparaenum}[(i)] 
	\item Nodal nontrivial superconductors display a noiseless zero-bias peak and their ratio is zero; 
	\item chiral superconductors feature a conductance of magnitude comparable to the normal state with noise-current ratio $1$; and
	\item trivial (gapful and nodal) superconductors have a suppressed conductance and ratio approaching $2$. 
\end{inparaenum}


\begin{figure*}
	\includegraphics[width=1.\textwidth]{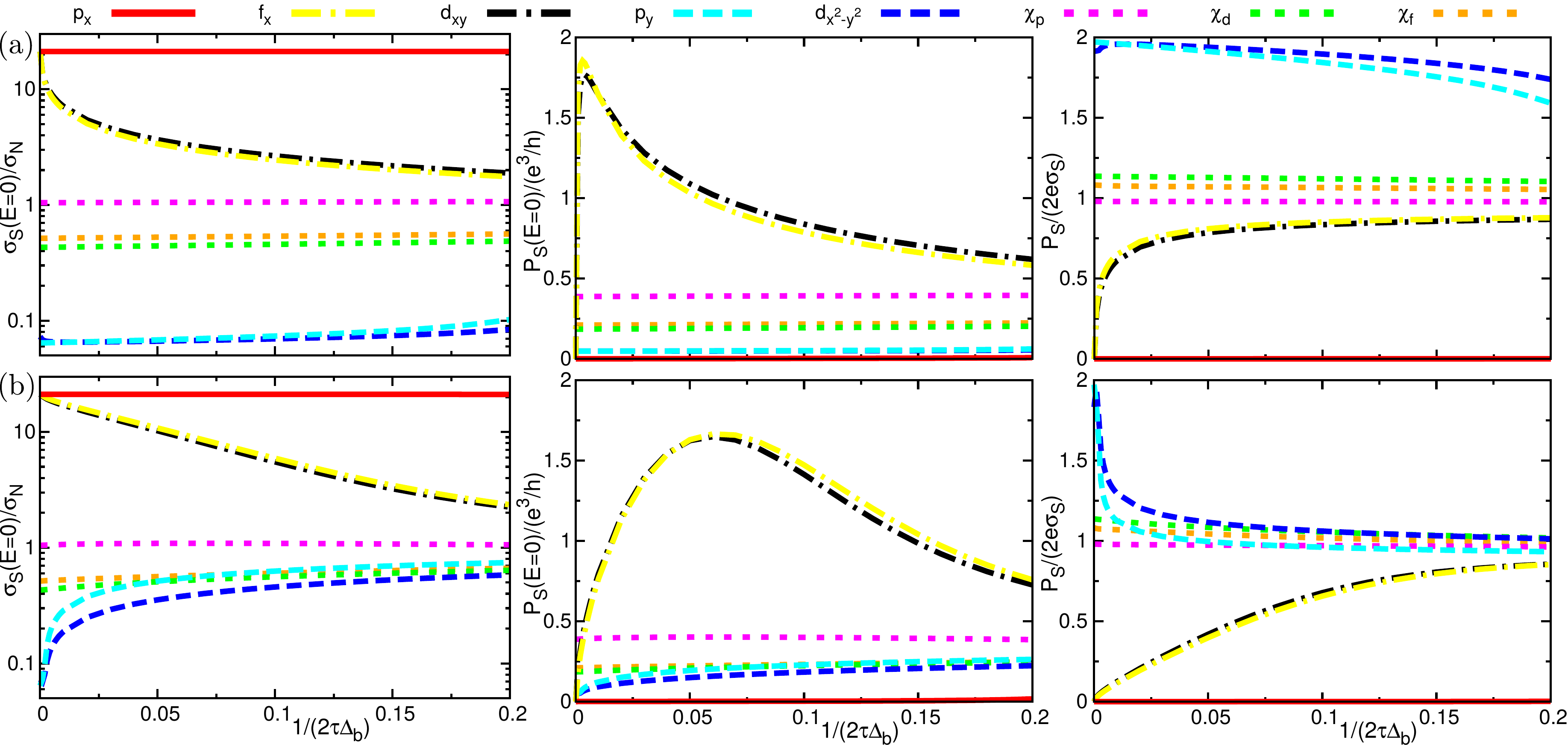}
	\caption{\label{fig:Fano_dis} 
		From left to right, differential conductance, noise power, and their ratio for $E\!=\!0$ as a function of the impurity scattering rate $1/\tau$ normalized to the bulk gap $\Delta_b$. In the left panels, the conductance is normalized to the normal state value. (a) Born limit with $\sigma\!=\!0$. (b) Unitary limit with $\sigma\!=\!0.99$. In all cases, $Z\!=\!5$. 
	}
\end{figure*}


\section{Impurity scattering in the superconductor \label{sec:imp}} 
We now consider the presence of non-magnetic impurities in the superconductor. We explore two cases for the impurity potential. The Born limit accounts for weak impurity potentials that induce small scattering phase shifts. We thus take the limit $\sigma\!\rightarrow\!0$ in \eref{eq:self-energy} and find
\begin{subequations}\label{eq:imp-born-unit}
\begin{equation}
\hat{a}\sim -\frac{1}{2\tau} \mean{ \mean{ \sum_j g_{j}^{\alpha\alpha} \hat{\tau}_{j}} }_{\theta} .
\end{equation}
On the other hand, the Unitary limit considers infinitely strong impurity potentials. Taking $\sigma\!\rightarrow\!1$ in \eref{eq:self-energy} results in
\begin{equation}
\hat{a}\sim \frac{1}{2\tau} \frac{1}{ \mean{ \mean{ \sum_j g_{j}^{\alpha\alpha} \hat{\tau}_{j} } }_{\theta} }.
\end{equation}
\end{subequations}
In the bulk of the superconductor, the Green's function becomes divergent for energies close to the continuum levels at the edges of the gap. Close to the interface, however, the Green's function can develop divergences in the presence of emergent SABS and it is approximately zero otherwise. Whether the superconductor is nontrivial and develops SABS or it is trivial and does not have subgap states crucially determines the impact of the impurity scattering\cite{Lu_2016}. 
For nontrivial superconductors, given a fixed scattering rate $1/\tau$, the self-energy $\hat{a}$ diverges in the Born limit and is greatly suppressed in the Unitary limit. For trivial superconductors we expect the opposite behavior. 

In the following, we set the barrier strength $Z\!=\!5$, where the general behavior of trivial and nontrivial unconventional superconducting orders is clearly displayed, and study the effect of scattering by non-magnetic impurities in the superconductor for zero and finite applied bias. 


\begin{figure*}
	\includegraphics[width=1.\textwidth]{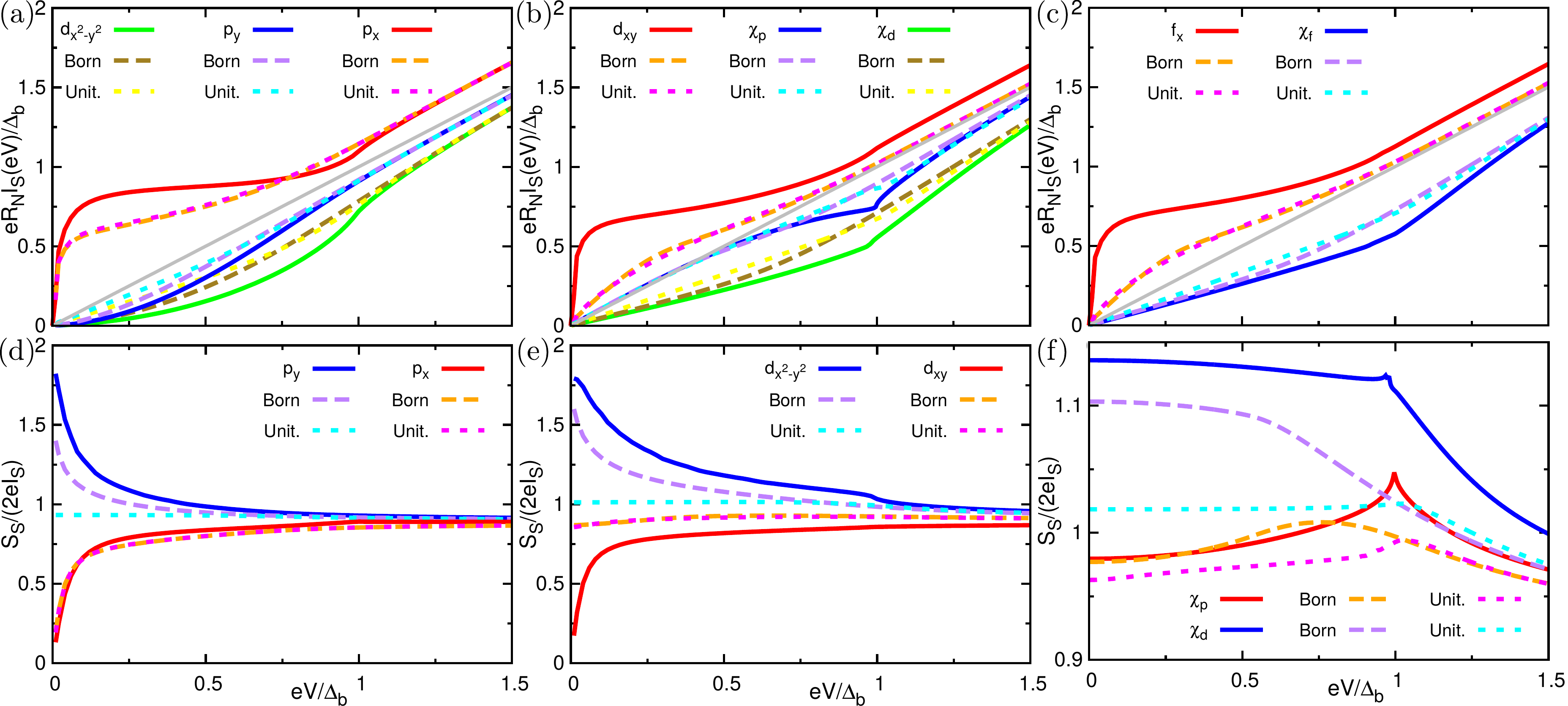}
	\caption{\label{fig:ivc} 
		Voltage dependence of the current (a-c) and noise-current ratio (d-f) for different pairing symmetries in the ballistic (solid lines), Born (dashed lines), and Unitary (dotted lines) limits. The latter two are plotted for $1/(2\tau\Delta_b)=0.2$. The gray line is normal state current. }
\end{figure*}


\subsection{Effect of impurities at zero bias. }
The zero bias transport results, for a tunnel junction with $Z\!=\!5$, are shown in \fref{fig:Fano_dis}(a) for the Born limit and in \fref{fig:Fano_dis}(b) for the Unitary one. 
In both cases, we immediately observe a different behavior between $p_x$-wave and the rest of nodal nontrivial states. 
For the moderate impurity scattering rates considered in this work, $p_x$-wave superconductors are immune to the effect of non-magnetic impurities. The ballistic zero-bias noiseless conductance is unaltered in Born and Unitary limits. Conversely, for $d_{xy}$- and $f_x$-wave cases the zero-bias conductance peak is reduced by the impurity potential. As the impurity strength is increased, conductance and shot noise tend to the same value, with ratio equal to $1$. 
It is interesting to note that the noise power for these superconductors, which is zero in the ballistic limit, develops a maximum as a function of the scattering rate $1/\tau$, similarly to the barrier dependence in the ballistic limit for the rest of superconductors. 
Nodal trivial superconductors, like $p_y$- and $d_{x^2-y^2}$-wave cases, are also strongly modified by impurities. In the Unitary limit, the conductance and shot noise of these superconductors are increased at zero bias, in clear tendency toward the normal state case. A similar trend is observed in the Born limit, although the evolution with the scattering rate is very smooth. 
By looking at the noise-current ratio, this tendency becomes evident for both Born and Unitary limits (rightmost panels of \fref{fig:Fano_dis}). 
The ratio is reduced from $2$ to $1$ for trivial pairings, and increased from $0$ to $1$ for nontrivial ones, with the notable exception of $p_x$-wave. 
As expected from \eref{eq:imp-born-unit}, this transition is faster in the Born limit than in the Unitary one for nontrivial superconductors, while trivial pairings follow the opposite behavior. 

The special case of $p_x$-wave pairing can be explained by the  emergence of isotropic odd-frequency Cooper pairs at the interface\cite{Tanaka_2007,Tanaka_2007c}. 
Inhomogeneous superconducting systems feature an ubiquitous presence of odd-frequency pairing\cite{Bergeret_2001,*Bergeret_RMP,*Linder_2010c,*Linder_2015,*Linder_2015b,Tanaka_2007b,Yokoyama_2012,*Black-Schaffer_2012,*Crepin_2015,*Burset_2015,*Aperis_2015,Sothmann_2014,*Burset_2016,Black-Schaffer_2013b,*Asano_2015}, where the wave function of Cooper pairs is odd under the exchange of the time coordinates of the electrons forming it\cite{Tanaka_JPSJ,Eschrig_RPP,Mizushima_2015}. 
In the ballistic limit, the noiseless zero-bias conductance peak of nodal nontrivial superconductors is formed due to a dominant odd-frequency pairing state near the interface\cite{Lu_2016}. 
To fulfill Fermi-Dirac statistics, these Cooper pairs form a $s$-wave state for $p_x$-wave superconductors or a $p$-wave state ($d$-wave) for $d_{xy}$-wave ($f_{x}$-wave) superconductors. The presence of impurities suppresses all the anisotropic pairing states and only $p_x$-wave order maintains the ballistic result. 

Finally, chiral superconductors are mostly unaffected by the impurity scattering. This behavior is clearly shown by the noise-current ratio, which remains almost $1$ in both Born and Unitary limits for several values of the scattering rate. 
At first sight this result seems at odds with the fact that nontrivial superconductors should be sensitive to impurity scattering in the Born limit. The main difference is that gapful nontrivial superconductors feature SABS with a linear dispersion relation, instead of flat bands like nodal superconductors. The resulting angle-averaged Green's function at low energies for chiral superconductors is not divergent, since the SABS only contribute for specific values of the angle (e.g., for $E\!=\!0$, the chiral $p$-wave SABS has a small contribution at $|\theta|\!\sim\!0$). 
This fundamental difference justifies the importance of chiral superconductors as sources of SABS with topological protection against disorder\cite{Chiral_SC_2016,Beenakker_2015}. 

\subsection{Effect of impurities at finite bias. }
We now study the voltage dependence of the current and fluctuations. 
In the normal state, the current through the junction follows a linear ohmic behavior where $I_N\!=\!R_N^{-1}V$, with $R_N$ the normal state resistance and $V$ the applied voltage (see gray lines in \fref{fig:ivc}). 
In the superconducting state, the current is drastically changed for bias values comparable to the superconducting gap $\Delta_b$. 
The $I\!-\!V$ curves of conventional singlet $s$-wave normal-metal--superconductor junctions in the ballistic limit are well-known, see, e.g., Refs.~\onlinecite{BTK,Cuevas_1996}. In the tunnel regime, the current is suppressed for voltages below the gap, while it is linear with slope $R_N^{-1}$ for high voltages $eV\!>\!\Delta_b$. 
This conventional behavior is qualitatively reproduced by nodal trivial superconductors with $p_y$- and $d_{x^2-y^2}$-wave symmetries. The main difference being a less pronounced suppression below the gap in the ballistic case; see solid blue and green lines of \fref{fig:ivc}(a). In the presence of impurities, the subgap suppression is even milder (dashed lines for the Born limit and dotted lines for the Unitary limit). 

Chiral $p$-wave superconductors, blue line in \fref{fig:ivc}(b), are clearly distinguished from chiral $d$- and $f$-wave cases, green line in \fref{fig:ivc}(b) and blue line in \fref{fig:ivc}(c), respectively. 
While chiral $p$-wave superconductors mostly follow an ohmic behavior, with a small dip at $eV\!\sim\!\Delta_b$, the current for the other chiral waves is suppressed below $I_N$, even for voltages over the gap. 
The finite voltage current is thus helpful to distinguish chiral $p$-wave symmetry from $d$-wave or higher, which is qualitatively similar to trivial superconductors. 
The main effect of impurity scattering on chiral superconductors is to soften the dip at $eV\!\sim\!\Delta_b$, where the presence of continuum bands makes $\mean{\mean{\sum_{j} \hat{g}_{j}(E\!\sim\!\Delta_b,x\!=\!0) \hat{\tau}_{j} }}_{\theta}$ diverge\cite{Bakurskiy_2014,Lu_2016}. As a consequence, the self-energy is acutely increased in the Born limit, while it is suppressed in the unitary limit; c.f., \eref{eq:imp-born-unit}. 
The current dips, which are clearly distinguishable in the ballistic limit, can still be appreciated in the Unitary limit but are completely suppressed in the Born limit. 

Nodal nontrivial superconductors display a completely different behavior. Even in the tunnel regime considered here with $Z\!=\!5$, the subgap current is greatly enhanced in the ballistic limit. The impurity scattering reduces the subgap contribution of $d_{xy}$- and $f_x$-wave superconductors, although the current is still greater than $I_N$. Conversely, the current for $p_x$-wave superconductors is enhanced over the ohmic case even in the presence of impurities. 

As for the zero-bias results, the noise-current ratio clearly displays the behavior of each pairing state, as it is shown in \fref{fig:ivc}(d-f)\footnotemark[3]. The ohmic $I\!-\!V$ curves for $eV\!>\!\Delta_b$ yield a ratio $1$. For nodal trivial gaps, the ballistic, Born and Unitary limits present different ratios going from $2$ (ballistic) to $1$ (Unitary). 
Nodal nontrivial pairings quickly approach the ohmic limit in the presence of impurities, with the exception of $p_x$-wave state. 
In \fref{fig:ivc}(f) we compare the chiral $p$-wave and $d$-wave states (which also qualitatively represents chiral $f$-wave). Chiral $p$-wave is only slightly modified by impurities at finite voltage. At low voltage, the ratio is only slightly affected in the Unitary limit. However, the rest of chiral states are a bit more sensitive, although in a small scale. 

\footnotetext[3]{The noise-current ratio in \fref{fig:ivc}(d-f) is only well defined for finite voltage, since $I_S(V\!=\!0)\!=\!0$. }


\begin{figure}
	\includegraphics[width=1.\columnwidth]{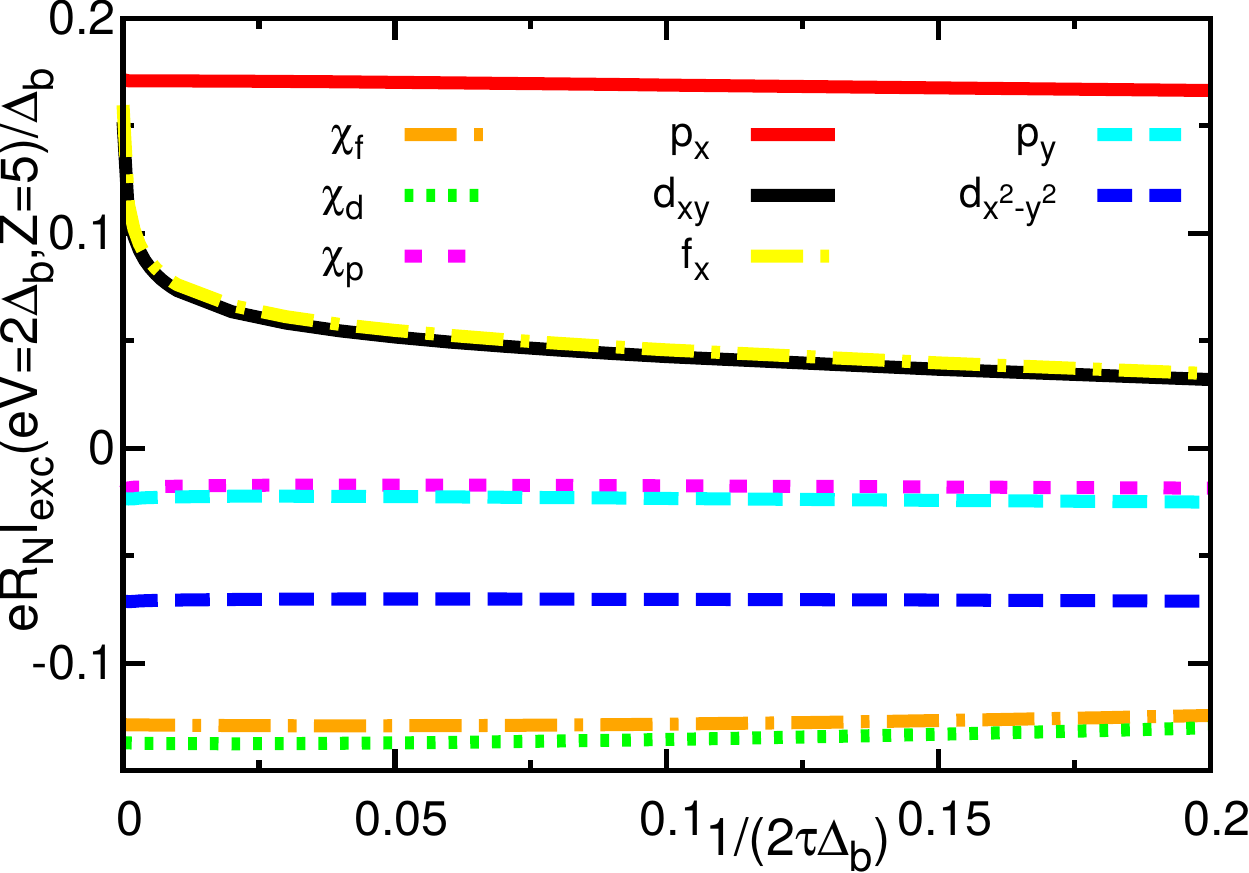}
	\caption{\label{fig:iexc} 
		In the Born limit with $1/(2\tau\Delta_b)=0.2$, excess current for $eV\!=\!2\Delta_b$ as a function of the impurity scattering rate. }
\end{figure}


In order to study the $I\!-\!V$ characteristics at high voltages, we define the excess current as the difference between the superconducting and normal state currents, namely,
\begin{equation}\label{eq:exc}
I_{\mathrm{exc}}(V)= I_S(eV)-I_{N}(eV) .
\end{equation}
The presence of a finite excess current for high voltages, ideally for $eV\!\rightarrow\!\infty$, indicates a strong contribution of Andreev reflections for energies below the gap\cite{BTK,Cuevas_1996,Scheer_2001,Jonas_2016}. 
In the ballistic limit, nodal nontrivial superconductors feature a finite excess current while it is suppressed for the rest of superconductors at large voltages. 
The value of the maximum excess current is determined by the interface transparency, resulting in $I_{\mathrm{exc}}^{\mathrm{ball}}(Z\!=\!5)\!\simeq\!(1/6)\Delta_b/(eR_N)$. 
In \fref{fig:iexc}, we show the evolution of the excess current as a function of the impurity scattering rate, calculated at $eV\!=\!2\Delta_b$. Negative values of $I_{\mathrm{exc}}(2\Delta_b)$ indicate that the excess current is suppressed for $eV\!\gg\!\Delta_b$. 
It is interesting that the presence of impurity scattering does not accelerate the transition into $I_N$ for chiral or trivial superconductors. 
However, impurity scattering suppresses the excess current for nodal nontrivial superconductors, with the exception of $p_x$-wave. 
Remarkably, $p_x$-wave superconductors maintain the ballistic result even in the presence of impurities, even though the excess current for the rest of nodal nontrivial cases is reduced to the normal state case. 
The presence of a noiseless perfect Andreev reflection in the zero energy channel of $p_x$-wave superconductors is thus directly responsible for a finite excess current, even in the presence of impurities. 


\section{Conclusions \label{sec:conc}}
We present an exhaustive description of the transport properties of two-dimensional junctions with unconventional superconductors, including the effect of scattering by non-magnetic impurities in the superconductor. 
The main results of this work are gathered in Table~\ref{table}. 
We have classified two-dimensional superconducting order parameters as gapful or nodal, where the latter vanishes for a particular direction of the wave vector. Each class can be topologically nontrivial if the superconductor features SABS. 
The noise-current ratio is a perfect tool to identify each class in an impurity-free ballistic junction in the tunnel limit. Indeed, the ratio is zero for nodal nontrivial superconductors, $1$ for gapful nontrivial ones, and $2$ for trivial pairings, both nodal or gapful. 

The inclusion of impurity scattering at the superconductor further distinguishes unconventional superconductors, and these changes are again clearly captured in the noise-current ratio. The ratio for trivial superconductors is decreased from $2$ to $1$ as the scattering rate is increased. This transition is faster in the Unitary limit; dominant in the absence of SABS. 
Conversely, the ratio for nodal nontrivial superconductors is increased from $0$ to $1$, and the transition is faster in the Born limit. 
A notable exception is the case of $p_x$-wave. This special nodal pairing develops an $s$-wave odd-frequency component at the interface which is highly resistant to impurity scattering. 
Interestingly, gapful nontrivial (chiral) superconductors are also resistant to the impurity scattering. The linear dispersion of the SABS in these superconductors guarantees that the correction coming from the self-energy of the impurity distribution is always small. Their noise-current ratio is thus barely changed by the presence of non-magnetic impurities, a clear signature of the topological protection of these pairing states. 

Our results can be used for the experimental classification of pairing symmetries. We have demonstrated the utility of noise measurements for the identification of order parameters, even in the cases where tunnel conductance can be ambiguous\cite{Brinkman_2016}. 

A special mention should be made about $p_x$-wave pairing. 
The flat band SABS of this nodal superconductor induces a perfect Andreev reflection, resulting in an noiseless resonance at zero energy. This behavior survives the presence of non-magnetic impurities in the superconductor and can be observed even when the normal metal is in the diffusive regime\cite{Kashiwaya_2004,*Yokoyama_2005}. 
The origin of this anomalous proximity effect is the formation of isotropic odd-frequency Cooper pairs at the interface\cite{Tanaka_2007}. 
Another manifestation of such an exotic odd-frequency Andreev resonance presented here is that the excess current maintains the ballistic maximum value in the presence of impurities; a feature unique to this pairing state. 
The special zero-energy state of $p_x$-wave superconductors is connected to the emergence of a Majorana bound state\cite{Ikegaya_2016}. Indeed, Majorana states are always accompanied by odd-frequency Cooper pairs\cite{Asano_2013,Kashuba_2017}. 
It is thus very motivating that a finite excess current has been recently reported in experiments to identify Majorana bound states in topological Josephson junctions\cite{Jonas_2016}. 

Finally, the connection between symmetry of the pairing state and transport properties of two-dimensional spin-degenerate superconducting junctions presented here provides a good starting point to consider more complicated three-dimensional pairing states. 


\acknowledgments 
The authors are grateful to Y. Asano for valuable discussions. 
P.B. acknowledges support from Japan Society for the Promotion of Science International Research Fellowship. 
B.L. acknowledges EPSRC grant EP/N010345/1 and European Graphene Flagship Project. 
This work was supported by a Grant-in-Aid for Scientific Research on Innovative Areas Topological Material Science JPSJ KAKENHI (Grants No. JP15H05851, and No. JP15H05853), a Grant-in-Aid for Scientific Research B (Grant No. JP15H03686), a Grant-in-Aid for Challenging Exploratory Research (Grant No. JP15K13498) from the Ministry of Education, Culture, Sports, Science, and Technology, Japan (MEXT). 






%


\begin{thebibliography}{88}%
	\makeatletter
	\providecommand \@ifxundefined [1]{%
		\@ifx{#1\undefined}
	}%
	\providecommand \@ifnum [1]{%
		\ifnum #1\expandafter \@firstoftwo
		\else \expandafter \@secondoftwo
		\fi
	}%
	\providecommand \@ifx [1]{%
		\ifx #1\expandafter \@firstoftwo
		\else \expandafter \@secondoftwo
		\fi
	}%
	\providecommand \natexlab [1]{#1}%
	\providecommand \enquote  [1]{``#1''}%
	\providecommand \bibnamefont  [1]{#1}%
	\providecommand \bibfnamefont [1]{#1}%
	\providecommand \citenamefont [1]{#1}%
	\providecommand \href@noop [0]{\@secondoftwo}%
	\providecommand \href [0]{\begingroup \@sanitize@url \@href}%
	\providecommand \@href[1]{\@@startlink{#1}\@@href}%
	\providecommand \@@href[1]{\endgroup#1\@@endlink}%
	\providecommand \@sanitize@url [0]{\catcode `\\12\catcode `\$12\catcode
		`\&12\catcode `\#12\catcode `\^12\catcode `\_12\catcode `\%12\relax}%
	\providecommand \@@startlink[1]{}%
	\providecommand \@@endlink[0]{}%
	\providecommand \url  [0]{\begingroup\@sanitize@url \@url }%
	\providecommand \@url [1]{\endgroup\@href {#1}{\urlprefix }}%
	\providecommand \urlprefix  [0]{URL }%
	\providecommand \Eprint [0]{\href }%
	\providecommand \doibase [0]{http://dx.doi.org/}%
	\providecommand \selectlanguage [0]{\@gobble}%
	\providecommand \bibinfo  [0]{\@secondoftwo}%
	\providecommand \bibfield  [0]{\@secondoftwo}%
	\providecommand \translation [1]{[#1]}%
	\providecommand \BibitemOpen [0]{}%
	\providecommand \bibitemStop [0]{}%
	\providecommand \bibitemNoStop [0]{.\EOS\space}%
	\providecommand \EOS [0]{\spacefactor3000\relax}%
	\providecommand \BibitemShut  [1]{\csname bibitem#1\endcsname}%
	\let\auto@bib@innerbib\@empty
	\bibitem [{\citenamefont {Sigrist}\ and\ \citenamefont
		{Ueda}(1991)}]{Sigrist_RMP}%
	\BibitemOpen
	\bibfield  {author} {\bibinfo {author} {\bibfnamefont {Manfred}\ \bibnamefont
			{Sigrist}}\ and\ \bibinfo {author} {\bibfnamefont {Kazuo}\ \bibnamefont
			{Ueda}},\ }\bibfield  {title} {\enquote {\bibinfo {title} {Phenomenological
				theory of unconventional superconductivity},}\ }\href {\doibase
		10.1103/RevModPhys.63.239} {\bibfield  {journal} {\bibinfo  {journal} {Rev.
				Mod. Phys.}\ }\textbf {\bibinfo {volume} {63}},\ \bibinfo {pages} {239--311}
		(\bibinfo {year} {1991})}\BibitemShut {NoStop}%
	\bibitem [{\citenamefont {Buchholtz}\ and\ \citenamefont
		{Zwicknagl}(1981)}]{Zwicknagl_1981}%
	\BibitemOpen
	\bibfield  {author} {\bibinfo {author} {\bibfnamefont {L.~J.}\ \bibnamefont
			{Buchholtz}}\ and\ \bibinfo {author} {\bibfnamefont {G.}~\bibnamefont
			{Zwicknagl}},\ }\bibfield  {title} {\enquote {\bibinfo {title}
			{Identification of $p$-wave superconductors},}\ }\href {\doibase
		10.1103/PhysRevB.23.5788} {\bibfield  {journal} {\bibinfo  {journal} {Phys.
				Rev. B}\ }\textbf {\bibinfo {volume} {23}},\ \bibinfo {pages} {5788--5796}
		(\bibinfo {year} {1981})}\BibitemShut {NoStop}%
	\bibitem [{\citenamefont {Bruder}(1990)}]{Bruder_1990}%
	\BibitemOpen
	\bibfield  {author} {\bibinfo {author} {\bibfnamefont {C.}~\bibnamefont
			{Bruder}},\ }\bibfield  {title} {\enquote {\bibinfo {title} {Andreev
				scattering in anisotropic superconductors},}\ }\href {\doibase
		10.1103/PhysRevB.41.4017} {\bibfield  {journal} {\bibinfo  {journal} {Phys.
				Rev. B}\ }\textbf {\bibinfo {volume} {41}},\ \bibinfo {pages} {4017--4032}
		(\bibinfo {year} {1990})}\BibitemShut {NoStop}%
	\bibitem [{\citenamefont {Hu}(1994)}]{Hu_1994}%
	\BibitemOpen
	\bibfield  {author} {\bibinfo {author} {\bibfnamefont {Chia-Ren}\
			\bibnamefont {Hu}},\ }\bibfield  {title} {\enquote {\bibinfo {title} {Midgap
				surface states as a novel signature for
				${\mathit{d}}_{\mathit{x}\mathit{a}}^{2}$-${\mathit{x}}_{\mathit{b}}^{2}$-wave
				superconductivity},}\ }\href {\doibase 10.1103/PhysRevLett.72.1526}
	{\bibfield  {journal} {\bibinfo  {journal} {Phys. Rev. Lett.}\ }\textbf
		{\bibinfo {volume} {72}},\ \bibinfo {pages} {1526--1529} (\bibinfo {year}
		{1994})}\BibitemShut {NoStop}%
	\bibitem [{\citenamefont {Tanaka}\ and\ \citenamefont
		{Kashiwaya}(1995)}]{Tanaka_1995}%
	\BibitemOpen
	\bibfield  {author} {\bibinfo {author} {\bibfnamefont {Yukio}\ \bibnamefont
			{Tanaka}}\ and\ \bibinfo {author} {\bibfnamefont {Satoshi}\ \bibnamefont
			{Kashiwaya}},\ }\bibfield  {title} {\enquote {\bibinfo {title} {Theory of
				tunneling spectroscopy of $\mathit{d}$-wave superconductors},}\ }\href
	{\doibase 10.1103/PhysRevLett.74.3451} {\bibfield  {journal} {\bibinfo
			{journal} {Phys. Rev. Lett.}\ }\textbf {\bibinfo {volume} {74}},\ \bibinfo
		{pages} {3451--3454} (\bibinfo {year} {1995})}\BibitemShut {NoStop}%
	\bibitem [{\citenamefont {Kashiwaya}\ and\ \citenamefont
		{Tanaka}(2000)}]{Kashiwaya_2000}%
	\BibitemOpen
	\bibfield  {author} {\bibinfo {author} {\bibfnamefont {Satoshi}\ \bibnamefont
			{Kashiwaya}}\ and\ \bibinfo {author} {\bibfnamefont {Yukio}\ \bibnamefont
			{Tanaka}},\ }\bibfield  {title} {\enquote {\bibinfo {title} {Tunnelling
				effects on surface bound states in unconventional superconductors},}\ }\href
	{\doibase 10.1088/0034-4885/63/10/202} {\bibfield  {journal} {\bibinfo
			{journal} {Reports on Progress in Physics}\ }\textbf {\bibinfo {volume}
			{63}},\ \bibinfo {pages} {1641} (\bibinfo {year} {2000})}\BibitemShut
	{NoStop}%
	\bibitem [{\citenamefont {Anderson}(1959)}]{Anderson_1959}%
	\BibitemOpen
	\bibfield  {author} {\bibinfo {author} {\bibfnamefont {P.W.}\ \bibnamefont
			{Anderson}},\ }\bibfield  {title} {\enquote {\bibinfo {title} {Theory of
				dirty superconductors},}\ }\href {\doibase
		http://dx.doi.org/10.1016/0022-3697(59)90036-8} {\bibfield  {journal}
		{\bibinfo  {journal} {Journal of Physics and Chemistry of Solids}\ }\textbf
		{\bibinfo {volume} {11}},\ \bibinfo {pages} {26 -- 30} (\bibinfo {year}
		{1959})}\BibitemShut {NoStop}%
	\bibitem [{\citenamefont {Balatsky}\ \emph {et~al.}(2006)\citenamefont
		{Balatsky}, \citenamefont {Vekhter},\ and\ \citenamefont
		{Zhu}}]{Balatsky_RMP}%
	\BibitemOpen
	\bibfield  {author} {\bibinfo {author} {\bibfnamefont {A.~V.}\ \bibnamefont
			{Balatsky}}, \bibinfo {author} {\bibfnamefont {I.}~\bibnamefont {Vekhter}}, \
		and\ \bibinfo {author} {\bibfnamefont {Jian-Xin}\ \bibnamefont {Zhu}},\
	}\bibfield  {title} {\enquote {\bibinfo {title} {Impurity-induced states in
			conventional and unconventional superconductors},}\ }\href {\doibase
	10.1103/RevModPhys.78.373} {\bibfield  {journal} {\bibinfo  {journal} {Rev.
			Mod. Phys.}\ }\textbf {\bibinfo {volume} {78}},\ \bibinfo {pages} {373--433}
	(\bibinfo {year} {2006})}\BibitemShut {NoStop}%
\bibitem [{\citenamefont {Schnyder}\ \emph {et~al.}(2008)\citenamefont
	{Schnyder}, \citenamefont {Ryu}, \citenamefont {Furusaki},\ and\
	\citenamefont {Ludwig}}]{Schnyder_2008}%
\BibitemOpen
\bibfield  {author} {\bibinfo {author} {\bibfnamefont {Andreas~P.}\
		\bibnamefont {Schnyder}}, \bibinfo {author} {\bibfnamefont {Shinsei}\
		\bibnamefont {Ryu}}, \bibinfo {author} {\bibfnamefont {Akira}\ \bibnamefont
		{Furusaki}}, \ and\ \bibinfo {author} {\bibfnamefont {Andreas W.~W.}\
		\bibnamefont {Ludwig}},\ }\bibfield  {title} {\enquote {\bibinfo {title}
		{Classification of topological insulators and superconductors in three
			spatial dimensions},}\ }\href {\doibase 10.1103/PhysRevB.78.195125}
{\bibfield  {journal} {\bibinfo  {journal} {Phys. Rev. B}\ }\textbf {\bibinfo
		{volume} {78}},\ \bibinfo {pages} {195125} (\bibinfo {year}
	{2008})}\BibitemShut {NoStop}%
\bibitem [{\citenamefont {Sato}(2009)}]{Sato_2009}%
\BibitemOpen
\bibfield  {author} {\bibinfo {author} {\bibfnamefont {Masatoshi}\
		\bibnamefont {Sato}},\ }\bibfield  {title} {\enquote {\bibinfo {title}
		{Topological properties of spin-triplet superconductors and fermi surface
			topology in the normal state},}\ }\href {\doibase 10.1103/PhysRevB.79.214526}
{\bibfield  {journal} {\bibinfo  {journal} {Phys. Rev. B}\ }\textbf {\bibinfo
		{volume} {79}},\ \bibinfo {pages} {214526} (\bibinfo {year}
	{2009})}\BibitemShut {NoStop}%
\bibitem [{\citenamefont {Sato}(2010)}]{Sato_2010}%
\BibitemOpen
\bibfield  {author} {\bibinfo {author} {\bibfnamefont {Masatoshi}\
		\bibnamefont {Sato}},\ }\bibfield  {title} {\enquote {\bibinfo {title}
		{Topological odd-parity superconductors},}\ }\href {\doibase
	10.1103/PhysRevB.81.220504} {\bibfield  {journal} {\bibinfo  {journal} {Phys.
			Rev. B}\ }\textbf {\bibinfo {volume} {81}},\ \bibinfo {pages} {220504}
	(\bibinfo {year} {2010})}\BibitemShut {NoStop}%
\bibitem [{\citenamefont {Ryu}\ \emph {et~al.}(2010)\citenamefont {Ryu},
	\citenamefont {Schnyder}, \citenamefont {Furusaki},\ and\ \citenamefont
	{Ludwig}}]{Schnyder_2010}%
\BibitemOpen
\bibfield  {author} {\bibinfo {author} {\bibfnamefont {Shinsei}\ \bibnamefont
		{Ryu}}, \bibinfo {author} {\bibfnamefont {Andreas~P}\ \bibnamefont
		{Schnyder}}, \bibinfo {author} {\bibfnamefont {Akira}\ \bibnamefont
		{Furusaki}}, \ and\ \bibinfo {author} {\bibfnamefont {Andreas W~W}\
		\bibnamefont {Ludwig}},\ }\bibfield  {title} {\enquote {\bibinfo {title}
		{Topological insulators and superconductors: tenfold way and dimensional
			hierarchy},}\ }\href {http://stacks.iop.org/1367-2630/12/i=6/a=065010}
{\bibfield  {journal} {\bibinfo  {journal} {New Journal of Physics}\ }\textbf
	{\bibinfo {volume} {12}},\ \bibinfo {pages} {065010} (\bibinfo {year}
	{2010})}\BibitemShut {NoStop}%
\bibitem [{\citenamefont {Qi}\ and\ \citenamefont {Zhang}(2011)}]{Qi_RMP}%
\BibitemOpen
\bibfield  {author} {\bibinfo {author} {\bibfnamefont {Xiao-Liang}\
		\bibnamefont {Qi}}\ and\ \bibinfo {author} {\bibfnamefont {Shou-Cheng}\
		\bibnamefont {Zhang}},\ }\bibfield  {title} {\enquote {\bibinfo {title}
		{Topological insulators and superconductors},}\ }\href {\doibase
	10.1103/RevModPhys.83.1057} {\bibfield  {journal} {\bibinfo  {journal} {Rev.
			Mod. Phys.}\ }\textbf {\bibinfo {volume} {83}},\ \bibinfo {pages}
	{1057--1110} (\bibinfo {year} {2011})}\BibitemShut {NoStop}%
\bibitem [{\citenamefont {Mizushima}\ \emph {et~al.}(2015)\citenamefont
	{Mizushima}, \citenamefont {Tsutsumi}, \citenamefont {Sato},\ and\
	\citenamefont {Machida}}]{Mizushima_2015}%
\BibitemOpen
\bibfield  {author} {\bibinfo {author} {\bibfnamefont {Takeshi}\ \bibnamefont
		{Mizushima}}, \bibinfo {author} {\bibfnamefont {Yasumasa}\ \bibnamefont
		{Tsutsumi}}, \bibinfo {author} {\bibfnamefont {Masatoshi}\ \bibnamefont
		{Sato}}, \ and\ \bibinfo {author} {\bibfnamefont {Kazushige}\ \bibnamefont
		{Machida}},\ }\bibfield  {title} {\enquote {\bibinfo {title} {Symmetry
			protected topological superfluid 3 he-b},}\ }\href
{http://stacks.iop.org/0953-8984/27/i=11/a=113203} {\bibfield  {journal}
	{\bibinfo  {journal} {Journal of Physics: Condensed Matter}\ }\textbf
	{\bibinfo {volume} {27}},\ \bibinfo {pages} {113203} (\bibinfo {year}
	{2015})}\BibitemShut {NoStop}%
\bibitem [{\citenamefont {Lu}\ \emph {et~al.}(2016)\citenamefont {Lu},
	\citenamefont {Burset}, \citenamefont {Tanuma}, \citenamefont {Golubov},
	\citenamefont {Asano},\ and\ \citenamefont {Tanaka}}]{Lu_2016}%
\BibitemOpen
\bibfield  {author} {\bibinfo {author} {\bibfnamefont {Bo}~\bibnamefont
		{Lu}}, \bibinfo {author} {\bibfnamefont {Pablo}\ \bibnamefont {Burset}},
	\bibinfo {author} {\bibfnamefont {Yasunari}\ \bibnamefont {Tanuma}}, \bibinfo
	{author} {\bibfnamefont {Alexander~A.}\ \bibnamefont {Golubov}}, \bibinfo
	{author} {\bibfnamefont {Yasuhiro}\ \bibnamefont {Asano}}, \ and\ \bibinfo
	{author} {\bibfnamefont {Yukio}\ \bibnamefont {Tanaka}},\ }\bibfield  {title}
{\enquote {\bibinfo {title} {Influence of the impurity scattering on charge
			transport in unconventional superconductor junctions},}\ }\href {\doibase
	10.1103/PhysRevB.94.014504} {\bibfield  {journal} {\bibinfo  {journal} {Phys.
			Rev. B}\ }\textbf {\bibinfo {volume} {94}},\ \bibinfo {pages} {014504}
	(\bibinfo {year} {2016})}\BibitemShut {NoStop}%
\bibitem [{\citenamefont {Blanter}\ and\ \citenamefont
	{Büttiker}(2000)}]{Blanter_2000}%
\BibitemOpen
\bibfield  {author} {\bibinfo {author} {\bibfnamefont {Ya.M.}\ \bibnamefont
		{Blanter}}\ and\ \bibinfo {author} {\bibfnamefont {M.}~\bibnamefont
		{Büttiker}},\ }\bibfield  {title} {\enquote {\bibinfo {title} {Shot noise in
			mesoscopic conductors},}\ }\href {\doibase
	http://dx.doi.org/10.1016/S0370-1573(99)00123-4} {\bibfield  {journal}
	{\bibinfo  {journal} {Physics Reports}\ }\textbf {\bibinfo {volume} {336}},\
	\bibinfo {pages} {1 -- 166} (\bibinfo {year} {2000})}\BibitemShut {NoStop}%
\bibitem [{\citenamefont {Khlus}(1987)}]{Khlus_1987}%
\BibitemOpen
\bibfield  {author} {\bibinfo {author} {\bibfnamefont {VA}~\bibnamefont
		{Khlus}},\ }\bibfield  {title} {\enquote {\bibinfo {title} {Current and
			voltage fluctuations in microjunctions between normal metals and
			superconductors},}\ }\href@noop {} {\bibfield  {journal} {\bibinfo  {journal}
		{Zh. Eksp. Teor. Fiz}\ }\textbf {\bibinfo {volume} {93}},\ \bibinfo {pages}
	{2179} (\bibinfo {year} {1987})}\BibitemShut {NoStop}%
\bibitem [{\citenamefont {Muzykantskii}\ and\ \citenamefont
	{Khmelnitskii}(1994)}]{noiseNS_1994}%
\BibitemOpen
\bibfield  {author} {\bibinfo {author} {\bibfnamefont {B.~A.}\ \bibnamefont
		{Muzykantskii}}\ and\ \bibinfo {author} {\bibfnamefont {D.~E.}\ \bibnamefont
		{Khmelnitskii}},\ }\bibfield  {title} {\enquote {\bibinfo {title} {Quantum
			shot noise in a normal-metal\char21{}superconductor point contact},}\ }\href
{\doibase 10.1103/PhysRevB.50.3982} {\bibfield  {journal} {\bibinfo
		{journal} {Phys. Rev. B}\ }\textbf {\bibinfo {volume} {50}},\ \bibinfo
	{pages} {3982--3987} (\bibinfo {year} {1994})}\BibitemShut {NoStop}%
\bibitem [{\citenamefont {de~Jong}\ and\ \citenamefont
	{Beenakker}(1994)}]{Beenakker_1994}%
\BibitemOpen
\bibfield  {author} {\bibinfo {author} {\bibfnamefont {M.~J.~M.}\
		\bibnamefont {de~Jong}}\ and\ \bibinfo {author} {\bibfnamefont {C.~W.~J.}\
		\bibnamefont {Beenakker}},\ }\bibfield  {title} {\enquote {\bibinfo {title}
		{Doubled shot noise in disordered normal-metal\char21{}superconductor
			junctions},}\ }\href {\doibase 10.1103/PhysRevB.49.16070} {\bibfield
	{journal} {\bibinfo  {journal} {Phys. Rev. B}\ }\textbf {\bibinfo {volume}
		{49}},\ \bibinfo {pages} {16070--16073} (\bibinfo {year} {1994})}\BibitemShut
{NoStop}%
\bibitem [{\citenamefont {Anantram}\ and\ \citenamefont
	{Datta}(1996)}]{Anantram-Datta}%
\BibitemOpen
\bibfield  {author} {\bibinfo {author} {\bibfnamefont {M.~P.}\ \bibnamefont
		{Anantram}}\ and\ \bibinfo {author} {\bibfnamefont {S.}~\bibnamefont
		{Datta}},\ }\bibfield  {title} {\enquote {\bibinfo {title} {Current
			fluctuations in mesoscopic systems with andreev scattering},}\ }\href
{\doibase 10.1103/PhysRevB.53.16390} {\bibfield  {journal} {\bibinfo
		{journal} {Phys. Rev. B}\ }\textbf {\bibinfo {volume} {53}},\ \bibinfo
	{pages} {16390--16402} (\bibinfo {year} {1996})}\BibitemShut {NoStop}%
\bibitem [{\citenamefont {Dieleman}\ \emph {et~al.}(1997)\citenamefont
	{Dieleman}, \citenamefont {Bukkems}, \citenamefont {Klapwijk}, \citenamefont
	{Schicke},\ and\ \citenamefont {Gundlach}}]{Klapwijk_1997}%
\BibitemOpen
\bibfield  {author} {\bibinfo {author} {\bibfnamefont {P.}~\bibnamefont
		{Dieleman}}, \bibinfo {author} {\bibfnamefont {H.~G.}\ \bibnamefont
		{Bukkems}}, \bibinfo {author} {\bibfnamefont {T.~M.}\ \bibnamefont
		{Klapwijk}}, \bibinfo {author} {\bibfnamefont {M.}~\bibnamefont {Schicke}}, \
	and\ \bibinfo {author} {\bibfnamefont {K.~H.}\ \bibnamefont {Gundlach}},\
}\bibfield  {title} {\enquote {\bibinfo {title} {Observation of andreev
		reflection enhanced shot noise},}\ }\href {\doibase
10.1103/PhysRevLett.79.3486} {\bibfield  {journal} {\bibinfo  {journal}
	{Phys. Rev. Lett.}\ }\textbf {\bibinfo {volume} {79}},\ \bibinfo {pages}
{3486--3489} (\bibinfo {year} {1997})}\BibitemShut {NoStop}%
\bibitem [{\citenamefont {Kallin}\ and\ \citenamefont
	{Berlinsky}(2016)}]{Chiral_SC_2016}%
\BibitemOpen
\bibfield  {author} {\bibinfo {author} {\bibfnamefont {Catherine}\
		\bibnamefont {Kallin}}\ and\ \bibinfo {author} {\bibfnamefont {John}\
		\bibnamefont {Berlinsky}},\ }\bibfield  {title} {\enquote {\bibinfo {title}
		{Chiral superconductors},}\ }\href
{http://stacks.iop.org/0034-4885/79/i=5/a=054502} {\bibfield  {journal}
	{\bibinfo  {journal} {Reports on Progress in Physics}\ }\textbf {\bibinfo
		{volume} {79}},\ \bibinfo {pages} {054502} (\bibinfo {year}
	{2016})}\BibitemShut {NoStop}%
\bibitem [{\citenamefont {Mackenzie}\ and\ \citenamefont
	{Maeno}(2003)}]{Mackenzie_2003}%
\BibitemOpen
\bibfield  {author} {\bibinfo {author} {\bibfnamefont {Andrew~Peter}\
		\bibnamefont {Mackenzie}}\ and\ \bibinfo {author} {\bibfnamefont {Yoshiteru}\
		\bibnamefont {Maeno}},\ }\bibfield  {title} {\enquote {\bibinfo {title} {The
			superconductivity of ${\mathrm{sr}}_{2}{\mathrm{ruo}}_{4}$ and the physics of
			spin-triplet pairing},}\ }\href {\doibase 10.1103/RevModPhys.75.657}
{\bibfield  {journal} {\bibinfo  {journal} {Rev. Mod. Phys.}\ }\textbf
	{\bibinfo {volume} {75}},\ \bibinfo {pages} {657--712} (\bibinfo {year}
	{2003})}\BibitemShut {NoStop}%
\bibitem [{\citenamefont {Maeno}\ \emph {et~al.}(2012)\citenamefont {Maeno},
	\citenamefont {Kittaka}, \citenamefont {Nomura}, \citenamefont {Yonezawa},\
	and\ \citenamefont {Ishida}}]{Maeno_2012}%
\BibitemOpen
\bibfield  {author} {\bibinfo {author} {\bibfnamefont {Yoshiteru}\
		\bibnamefont {Maeno}}, \bibinfo {author} {\bibfnamefont {Shunichiro}\
		\bibnamefont {Kittaka}}, \bibinfo {author} {\bibfnamefont {Takuji}\
		\bibnamefont {Nomura}}, \bibinfo {author} {\bibfnamefont {Shingo}\
		\bibnamefont {Yonezawa}}, \ and\ \bibinfo {author} {\bibfnamefont {Kenji}\
		\bibnamefont {Ishida}},\ }\bibfield  {title} {\enquote {\bibinfo {title}
		{Evaluation of spin-triplet superconductivity in sr$_{2}$ruo$_{4}$},}\ }\href
{\doibase 10.1143/JPSJ.81.011009} {\bibfield  {journal} {\bibinfo  {journal}
		{Journal of the Physical Society of Japan}\ }\textbf {\bibinfo {volume}
		{81}},\ \bibinfo {pages} {011009} (\bibinfo {year} {2012})}\BibitemShut
{NoStop}%
\bibitem [{\citenamefont {Kallin}(2012)}]{Kallin_2012}%
\BibitemOpen
\bibfield  {author} {\bibinfo {author} {\bibfnamefont {Catherine}\
		\bibnamefont {Kallin}},\ }\bibfield  {title} {\enquote {\bibinfo {title}
		{Chiral p-wave order in sr 2 ruo 4},}\ }\href
{http://stacks.iop.org/0034-4885/75/i=4/a=042501} {\bibfield  {journal}
	{\bibinfo  {journal} {Reports on Progress in Physics}\ }\textbf {\bibinfo
		{volume} {75}},\ \bibinfo {pages} {042501} (\bibinfo {year}
	{2012})}\BibitemShut {NoStop}%
\bibitem [{\citenamefont {Huang}\ \emph {et~al.}(2014)\citenamefont {Huang},
	\citenamefont {Taylor},\ and\ \citenamefont {Kallin}}]{Kallin_2014}%
\BibitemOpen
\bibfield  {author} {\bibinfo {author} {\bibfnamefont {Wen}\ \bibnamefont
		{Huang}}, \bibinfo {author} {\bibfnamefont {Edward}\ \bibnamefont {Taylor}},
	\ and\ \bibinfo {author} {\bibfnamefont {Catherine}\ \bibnamefont {Kallin}},\
}\bibfield  {title} {\enquote {\bibinfo {title} {Vanishing edge currents in
		non-$p$-wave topological chiral superconductors},}\ }\href {\doibase
10.1103/PhysRevB.90.224519} {\bibfield  {journal} {\bibinfo  {journal} {Phys.
		Rev. B}\ }\textbf {\bibinfo {volume} {90}},\ \bibinfo {pages} {224519}
(\bibinfo {year} {2014})}\BibitemShut {NoStop}%
\bibitem [{\citenamefont {Tada}\ \emph {et~al.}(2015)\citenamefont {Tada},
	\citenamefont {Nie},\ and\ \citenamefont {Oshikawa}}]{Masaki_2015}%
\BibitemOpen
\bibfield  {author} {\bibinfo {author} {\bibfnamefont {Yasuhiro}\
		\bibnamefont {Tada}}, \bibinfo {author} {\bibfnamefont {Wenxing}\
		\bibnamefont {Nie}}, \ and\ \bibinfo {author} {\bibfnamefont {Masaki}\
		\bibnamefont {Oshikawa}},\ }\bibfield  {title} {\enquote {\bibinfo {title}
		{Orbital angular momentum and spectral flow in two-dimensional chiral
			superfluids},}\ }\href {\doibase 10.1103/PhysRevLett.114.195301} {\bibfield
	{journal} {\bibinfo  {journal} {Phys. Rev. Lett.}\ }\textbf {\bibinfo
		{volume} {114}},\ \bibinfo {pages} {195301} (\bibinfo {year}
	{2015})}\BibitemShut {NoStop}%
\bibitem [{\citenamefont {Scaffidi}\ and\ \citenamefont
	{Simon}(2015)}]{Simon_2015}%
\BibitemOpen
\bibfield  {author} {\bibinfo {author} {\bibfnamefont {Thomas}\ \bibnamefont
		{Scaffidi}}\ and\ \bibinfo {author} {\bibfnamefont {Steven~H.}\ \bibnamefont
		{Simon}},\ }\bibfield  {title} {\enquote {\bibinfo {title} {Large chern
			number and edge currents in ${\mathrm{sr}}_{2}{\mathrm{ruo}}_{4}$},}\ }\href
{\doibase 10.1103/PhysRevLett.115.087003} {\bibfield  {journal} {\bibinfo
		{journal} {Phys. Rev. Lett.}\ }\textbf {\bibinfo {volume} {115}},\ \bibinfo
	{pages} {087003} (\bibinfo {year} {2015})}\BibitemShut {NoStop}%
\bibitem [{\citenamefont {Black-Schaffer}\ and\ \citenamefont
	{Honerkamp}(2014)}]{Chiral_graphene_2014}%
\BibitemOpen
\bibfield  {author} {\bibinfo {author} {\bibfnamefont {Annica~M}\
		\bibnamefont {Black-Schaffer}}\ and\ \bibinfo {author} {\bibfnamefont
		{Carsten}\ \bibnamefont {Honerkamp}},\ }\bibfield  {title} {\enquote
	{\bibinfo {title} {Chiral d -wave superconductivity in doped graphene},}\
}\href {http://stacks.iop.org/0953-8984/26/i=42/a=423201} {\bibfield
{journal} {\bibinfo  {journal} {Journal of Physics: Condensed Matter}\
}\textbf {\bibinfo {volume} {26}},\ \bibinfo {pages} {423201} (\bibinfo
{year} {2014})}\BibitemShut {NoStop}%
\bibitem [{\citenamefont {Read}\ and\ \citenamefont {Green}(2000)}]{Read_2000}%
\BibitemOpen
\bibfield  {author} {\bibinfo {author} {\bibfnamefont {N.}~\bibnamefont
		{Read}}\ and\ \bibinfo {author} {\bibfnamefont {Dmitry}\ \bibnamefont
		{Green}},\ }\bibfield  {title} {\enquote {\bibinfo {title} {Paired states of
			fermions in two dimensions with breaking of parity and time-reversal
			symmetries and the fractional quantum hall effect},}\ }\href {\doibase
	10.1103/PhysRevB.61.10267} {\bibfield  {journal} {\bibinfo  {journal} {Phys.
			Rev. B}\ }\textbf {\bibinfo {volume} {61}},\ \bibinfo {pages} {10267--10297}
	(\bibinfo {year} {2000})}\BibitemShut {NoStop}%
\bibitem [{\citenamefont {Fujimoto}(2008)}]{Fujimoto_2008}%
\BibitemOpen
\bibfield  {author} {\bibinfo {author} {\bibfnamefont {Satoshi}\ \bibnamefont
		{Fujimoto}},\ }\bibfield  {title} {\enquote {\bibinfo {title} {Topological
			order and non-abelian statistics in noncentrosymmetric $s$-wave
			superconductors},}\ }\href {\doibase 10.1103/PhysRevB.77.220501} {\bibfield
	{journal} {\bibinfo  {journal} {Phys. Rev. B}\ }\textbf {\bibinfo {volume}
		{77}},\ \bibinfo {pages} {220501} (\bibinfo {year} {2008})}\BibitemShut
{NoStop}%
\bibitem [{\citenamefont {Sau}\ \emph {et~al.}(2010)\citenamefont {Sau},
	\citenamefont {Lutchyn}, \citenamefont {Tewari},\ and\ \citenamefont
	{Das~Sarma}}]{Sau_2010}%
\BibitemOpen
\bibfield  {author} {\bibinfo {author} {\bibfnamefont {Jay~D.}\ \bibnamefont
		{Sau}}, \bibinfo {author} {\bibfnamefont {Roman~M.}\ \bibnamefont {Lutchyn}},
	\bibinfo {author} {\bibfnamefont {Sumanta}\ \bibnamefont {Tewari}}, \ and\
	\bibinfo {author} {\bibfnamefont {S.}~\bibnamefont {Das~Sarma}},\ }\bibfield
{title} {\enquote {\bibinfo {title} {Generic new platform for topological
			quantum computation using semiconductor heterostructures},}\ }\href {\doibase
	10.1103/PhysRevLett.104.040502} {\bibfield  {journal} {\bibinfo  {journal}
		{Phys. Rev. Lett.}\ }\textbf {\bibinfo {volume} {104}},\ \bibinfo {pages}
	{040502} (\bibinfo {year} {2010})}\BibitemShut {NoStop}%
\bibitem [{\citenamefont {Lutchyn}\ \emph {et~al.}(2010)\citenamefont
	{Lutchyn}, \citenamefont {Sau},\ and\ \citenamefont
	{Das~Sarma}}]{Lutchyn_2010}%
\BibitemOpen
\bibfield  {author} {\bibinfo {author} {\bibfnamefont {Roman~M.}\
		\bibnamefont {Lutchyn}}, \bibinfo {author} {\bibfnamefont {Jay~D.}\
		\bibnamefont {Sau}}, \ and\ \bibinfo {author} {\bibfnamefont
		{S.}~\bibnamefont {Das~Sarma}},\ }\bibfield  {title} {\enquote {\bibinfo
		{title} {Majorana fermions and a topological phase transition in
			semiconductor-superconductor heterostructures},}\ }\href {\doibase
	10.1103/PhysRevLett.105.077001} {\bibfield  {journal} {\bibinfo  {journal}
		{Phys. Rev. Lett.}\ }\textbf {\bibinfo {volume} {105}},\ \bibinfo {pages}
	{077001} (\bibinfo {year} {2010})}\BibitemShut {NoStop}%
\bibitem [{\citenamefont {Sato}\ \emph {et~al.}(2011)\citenamefont {Sato},
	\citenamefont {Tanaka}, \citenamefont {Yada},\ and\ \citenamefont
	{Yokoyama}}]{Yokoyama_2011}%
\BibitemOpen
\bibfield  {author} {\bibinfo {author} {\bibfnamefont {Masatoshi}\
		\bibnamefont {Sato}}, \bibinfo {author} {\bibfnamefont {Yukio}\ \bibnamefont
		{Tanaka}}, \bibinfo {author} {\bibfnamefont {Keiji}\ \bibnamefont {Yada}}, \
	and\ \bibinfo {author} {\bibfnamefont {Takehito}\ \bibnamefont {Yokoyama}},\
}\bibfield  {title} {\enquote {\bibinfo {title} {Topology of andreev bound
		states with flat dispersion},}\ }\href {\doibase 10.1103/PhysRevB.83.224511}
{\bibfield  {journal} {\bibinfo  {journal} {Phys. Rev. B}\ }\textbf {\bibinfo
		{volume} {83}},\ \bibinfo {pages} {224511} (\bibinfo {year}
	{2011})}\BibitemShut {NoStop}%
\bibitem [{\citenamefont {Kashiwaya}\ \emph {et~al.}(1995)\citenamefont
	{Kashiwaya}, \citenamefont {Tanaka}, \citenamefont {Koyanagi}, \citenamefont
	{Takashima},\ and\ \citenamefont {Kajimura}}]{Kashiwaya_1995}%
\BibitemOpen
\bibfield  {author} {\bibinfo {author} {\bibfnamefont {Satoshi}\ \bibnamefont
		{Kashiwaya}}, \bibinfo {author} {\bibfnamefont {Yukio}\ \bibnamefont
		{Tanaka}}, \bibinfo {author} {\bibfnamefont {Masao}\ \bibnamefont
		{Koyanagi}}, \bibinfo {author} {\bibfnamefont {Hiroshi}\ \bibnamefont
		{Takashima}}, \ and\ \bibinfo {author} {\bibfnamefont {Koji}\ \bibnamefont
		{Kajimura}},\ }\bibfield  {title} {\enquote {\bibinfo {title} {Origin of
			zero-bias conductance peaks in high-${\mathit{t}}_{\mathit{c}}$
			superconductors},}\ }\href {\doibase 10.1103/PhysRevB.51.1350} {\bibfield
	{journal} {\bibinfo  {journal} {Phys. Rev. B}\ }\textbf {\bibinfo {volume}
		{51}},\ \bibinfo {pages} {1350--1353} (\bibinfo {year} {1995})}\BibitemShut
{NoStop}%
\bibitem [{\citenamefont {Tanaka}\ \emph {et~al.}(2010)\citenamefont {Tanaka},
	\citenamefont {Mizuno}, \citenamefont {Yokoyama}, \citenamefont {Yada},\ and\
	\citenamefont {Sato}}]{Yada_2010}%
\BibitemOpen
\bibfield  {author} {\bibinfo {author} {\bibfnamefont {Yukio}\ \bibnamefont
		{Tanaka}}, \bibinfo {author} {\bibfnamefont {Yoshihiro}\ \bibnamefont
		{Mizuno}}, \bibinfo {author} {\bibfnamefont {Takehito}\ \bibnamefont
		{Yokoyama}}, \bibinfo {author} {\bibfnamefont {Keiji}\ \bibnamefont {Yada}},
	\ and\ \bibinfo {author} {\bibfnamefont {Masatoshi}\ \bibnamefont {Sato}},\
}\bibfield  {title} {\enquote {\bibinfo {title} {Anomalous andreev bound
		state in noncentrosymmetric superconductors},}\ }\href {\doibase
10.1103/PhysRevLett.105.097002} {\bibfield  {journal} {\bibinfo  {journal}
	{Phys. Rev. Lett.}\ }\textbf {\bibinfo {volume} {105}},\ \bibinfo {pages}
{097002} (\bibinfo {year} {2010})}\BibitemShut {NoStop}%
\bibitem [{\citenamefont {Yada}\ \emph {et~al.}(2011)\citenamefont {Yada},
	\citenamefont {Sato}, \citenamefont {Tanaka},\ and\ \citenamefont
	{Yokoyama}}]{Yada_2011}%
\BibitemOpen
\bibfield  {author} {\bibinfo {author} {\bibfnamefont {Keiji}\ \bibnamefont
		{Yada}}, \bibinfo {author} {\bibfnamefont {Masatoshi}\ \bibnamefont {Sato}},
	\bibinfo {author} {\bibfnamefont {Yukio}\ \bibnamefont {Tanaka}}, \ and\
	\bibinfo {author} {\bibfnamefont {Takehito}\ \bibnamefont {Yokoyama}},\
}\bibfield  {title} {\enquote {\bibinfo {title} {Surface density of states
		and topological edge states in noncentrosymmetric superconductors},}\ }\href
{\doibase 10.1103/PhysRevB.83.064505} {\bibfield  {journal} {\bibinfo
		{journal} {Phys. Rev. B}\ }\textbf {\bibinfo {volume} {83}},\ \bibinfo
	{pages} {064505} (\bibinfo {year} {2011})}\BibitemShut {NoStop}%
\bibitem [{\citenamefont {You}\ \emph {et~al.}(2013)\citenamefont {You},
	\citenamefont {Oh},\ and\ \citenamefont {Vedral}}]{Vedral_2013}%
\BibitemOpen
\bibfield  {author} {\bibinfo {author} {\bibfnamefont {Jiabin}\ \bibnamefont
		{You}}, \bibinfo {author} {\bibfnamefont {C.~H.}\ \bibnamefont {Oh}}, \ and\
	\bibinfo {author} {\bibfnamefont {Vlatko}\ \bibnamefont {Vedral}},\
}\bibfield  {title} {\enquote {\bibinfo {title} {Majorana fermions in
		$s$-wave noncentrosymmetric superconductor with dresselhaus (110) spin-orbit
		coupling},}\ }\href {\doibase 10.1103/PhysRevB.87.054501} {\bibfield
{journal} {\bibinfo  {journal} {Phys. Rev. B}\ }\textbf {\bibinfo {volume}
	{87}},\ \bibinfo {pages} {054501} (\bibinfo {year} {2013})}\BibitemShut
{NoStop}%
\bibitem [{\citenamefont {Ikegaya}\ \emph {et~al.}(2015)\citenamefont
	{Ikegaya}, \citenamefont {Asano},\ and\ \citenamefont
	{Tanaka}}]{Ikegaya_2015}%
\BibitemOpen
\bibfield  {author} {\bibinfo {author} {\bibfnamefont {Satoshi}\ \bibnamefont
		{Ikegaya}}, \bibinfo {author} {\bibfnamefont {Yasuhiro}\ \bibnamefont
		{Asano}}, \ and\ \bibinfo {author} {\bibfnamefont {Yukio}\ \bibnamefont
		{Tanaka}},\ }\bibfield  {title} {\enquote {\bibinfo {title} {Anomalous
			proximity effect and theoretical design for its realization},}\ }\href
{\doibase 10.1103/PhysRevB.91.174511} {\bibfield  {journal} {\bibinfo
		{journal} {Phys. Rev. B}\ }\textbf {\bibinfo {volume} {91}},\ \bibinfo
	{pages} {174511} (\bibinfo {year} {2015})}\BibitemShut {NoStop}%
\bibitem [{Note1()}]{Note1}%
\BibitemOpen
\bibinfo {note} {We are considering the situation where the nodal direction
	lies along the $x$-direction. In a more general case, there can be an angle
	between the nodal direction and the $x$-axis. A slightly tilted $p_y$-wave or
	$d_{x^2-y^2}$-wave pairing is then nontrivial, but it will behave in a
	similar way as the trivial cases studied here. We are only interested in the
	representative behavior for two-dimensional superconductors, so we will not
	consider such cases here.}\BibitemShut {Stop}%
\bibitem [{\citenamefont {Serene}\ and\ \citenamefont
	{Rainer}(1983)}]{Rainer_1983}%
\BibitemOpen
\bibfield  {author} {\bibinfo {author} {\bibfnamefont {J.W}\ \bibnamefont
		{Serene}}\ and\ \bibinfo {author} {\bibfnamefont {D}~\bibnamefont {Rainer}},\
}\bibfield  {title} {\enquote {\bibinfo {title} {The quasiclassical approach
		to superfluid 3he},}\ }\href {\doibase
http://dx.doi.org/10.1016/0370-1573(83)90051-0} {\bibfield  {journal}
{\bibinfo  {journal} {Physics Reports}\ }\textbf {\bibinfo {volume} {101}},\
\bibinfo {pages} {221 -- 311} (\bibinfo {year} {1983})}\BibitemShut {NoStop}%
\bibitem [{\citenamefont {Rammer}\ and\ \citenamefont
	{Smith}(1986)}]{Rammer_RMP}%
\BibitemOpen
\bibfield  {author} {\bibinfo {author} {\bibfnamefont {J.}~\bibnamefont
		{Rammer}}\ and\ \bibinfo {author} {\bibfnamefont {H.}~\bibnamefont {Smith}},\
}\bibfield  {title} {\enquote {\bibinfo {title} {Quantum field-theoretical
		methods in transport theory of metals},}\ }\href {\doibase
10.1103/RevModPhys.58.323} {\bibfield  {journal} {\bibinfo  {journal} {Rev.
		Mod. Phys.}\ }\textbf {\bibinfo {volume} {58}},\ \bibinfo {pages} {323--359}
(\bibinfo {year} {1986})}\BibitemShut {NoStop}%
\bibitem [{\citenamefont {Millis}\ \emph {et~al.}(1988)\citenamefont {Millis},
	\citenamefont {Rainer},\ and\ \citenamefont {Sauls}}]{Sauls_1988}%
\BibitemOpen
\bibfield  {author} {\bibinfo {author} {\bibfnamefont {A.}~\bibnamefont
		{Millis}}, \bibinfo {author} {\bibfnamefont {D.}~\bibnamefont {Rainer}}, \
	and\ \bibinfo {author} {\bibfnamefont {J.~A.}\ \bibnamefont {Sauls}},\
}\bibfield  {title} {\enquote {\bibinfo {title} {Quasiclassical theory of
		superconductivity near magnetically active interfaces},}\ }\href {\doibase
10.1103/PhysRevB.38.4504} {\bibfield  {journal} {\bibinfo  {journal} {Phys.
		Rev. B}\ }\textbf {\bibinfo {volume} {38}},\ \bibinfo {pages} {4504--4515}
(\bibinfo {year} {1988})}\BibitemShut {NoStop}%
\bibitem [{\citenamefont {Ashida}\ \emph {et~al.}(1989)\citenamefont {Ashida},
	\citenamefont {Aoyama}, \citenamefont {Hara},\ and\ \citenamefont
	{Nagai}}]{Nagai_1989}%
\BibitemOpen
\bibfield  {author} {\bibinfo {author} {\bibfnamefont {Masami}\ \bibnamefont
		{Ashida}}, \bibinfo {author} {\bibfnamefont {Shingo}\ \bibnamefont {Aoyama}},
	\bibinfo {author} {\bibfnamefont {Jun'ichiro}\ \bibnamefont {Hara}}, \ and\
	\bibinfo {author} {\bibfnamefont {Katsuhiko}\ \bibnamefont {Nagai}},\
}\bibfield  {title} {\enquote {\bibinfo {title} {Green's function in
		proximity-contact superconducting-normal double layers},}\ }\href {\doibase
10.1103/PhysRevB.40.8673} {\bibfield  {journal} {\bibinfo  {journal} {Phys.
		Rev. B}\ }\textbf {\bibinfo {volume} {40}},\ \bibinfo {pages} {8673--8686}
(\bibinfo {year} {1989})}\BibitemShut {NoStop}%
\bibitem [{\citenamefont {Eilenberger}(1968)}]{Eilenberger_1968}%
\BibitemOpen
\bibfield  {author} {\bibinfo {author} {\bibfnamefont {Gert}\ \bibnamefont
		{Eilenberger}},\ }\bibfield  {title} {\enquote {\bibinfo {title}
		{Transformation of gorkov's equation for type ii superconductors into
			transport-like equations},}\ }\href {\doibase 10.1007/BF01379803} {\bibfield
	{journal} {\bibinfo  {journal} {Zeitschrift f{\"u}r Physik A Hadrons and
			nuclei}\ }\textbf {\bibinfo {volume} {214}},\ \bibinfo {pages} {195--213}
	(\bibinfo {year} {1968})}\BibitemShut {NoStop}%
\bibitem [{\citenamefont {Tanaka}\ \emph
	{et~al.}(2007{\natexlab{a}})\citenamefont {Tanaka}, \citenamefont {Tanuma},\
	and\ \citenamefont {Golubov}}]{Tanaka_2007c}%
\BibitemOpen
\bibfield  {author} {\bibinfo {author} {\bibfnamefont {Y.}~\bibnamefont
		{Tanaka}}, \bibinfo {author} {\bibfnamefont {Y.}~\bibnamefont {Tanuma}}, \
	and\ \bibinfo {author} {\bibfnamefont {A.~A.}\ \bibnamefont {Golubov}},\
}\bibfield  {title} {\enquote {\bibinfo {title} {Odd-frequency pairing in
		normal-metal/superconductor junctions},}\ }\href {\doibase
10.1103/PhysRevB.76.054522} {\bibfield  {journal} {\bibinfo  {journal} {Phys.
		Rev. B}\ }\textbf {\bibinfo {volume} {76}},\ \bibinfo {pages} {054522}
(\bibinfo {year} {2007}{\natexlab{a}})}\BibitemShut {NoStop}%
\bibitem [{\citenamefont {Nagato}\ \emph {et~al.}(1993)\citenamefont {Nagato},
	\citenamefont {Nagai},\ and\ \citenamefont {Hara}}]{Nagato_1993}%
\BibitemOpen
\bibfield  {author} {\bibinfo {author} {\bibfnamefont {Yasushi}\ \bibnamefont
		{Nagato}}, \bibinfo {author} {\bibfnamefont {Katsuhiko}\ \bibnamefont
		{Nagai}}, \ and\ \bibinfo {author} {\bibfnamefont {Jun'ichiro}\ \bibnamefont
		{Hara}},\ }\bibfield  {title} {\enquote {\bibinfo {title} {Theory of the
			andreev reflection and the density of states in proximity contact
			normal-superconducting infinite double-layer},}\ }\href {\doibase
	10.1007/BF00682280} {\bibfield  {journal} {\bibinfo  {journal} {Journal of
			Low Temperature Physics}\ }\textbf {\bibinfo {volume} {93}},\ \bibinfo
	{pages} {33--56} (\bibinfo {year} {1993})}\BibitemShut {NoStop}%
\bibitem [{\citenamefont {Schopohl}\ and\ \citenamefont
	{Maki}(1995)}]{Kazumi_1995}%
\BibitemOpen
\bibfield  {author} {\bibinfo {author} {\bibfnamefont {Nils}\ \bibnamefont
		{Schopohl}}\ and\ \bibinfo {author} {\bibfnamefont {Kazumi}\ \bibnamefont
		{Maki}},\ }\bibfield  {title} {\enquote {\bibinfo {title} {Quasiparticle
			spectrum around a vortex line in a \textit{d}-wave superconductor},}\ }\href
{\doibase 10.1103/PhysRevB.52.490} {\bibfield  {journal} {\bibinfo  {journal}
		{Phys. Rev. B}\ }\textbf {\bibinfo {volume} {52}},\ \bibinfo {pages}
	{490--493} (\bibinfo {year} {1995})}\BibitemShut {NoStop}%
\bibitem [{\citenamefont {Shelankov}\ and\ \citenamefont
	{Ozana}(2000)}]{Ozana_2000}%
\BibitemOpen
\bibfield  {author} {\bibinfo {author} {\bibfnamefont {A.}~\bibnamefont
		{Shelankov}}\ and\ \bibinfo {author} {\bibfnamefont {M.}~\bibnamefont
		{Ozana}},\ }\bibfield  {title} {\enquote {\bibinfo {title} {Quasiclassical
			theory of superconductivity: A multiple-interface geometry},}\ }\href
{\doibase 10.1103/PhysRevB.61.7077} {\bibfield  {journal} {\bibinfo
		{journal} {Phys. Rev. B}\ }\textbf {\bibinfo {volume} {61}},\ \bibinfo
	{pages} {7077--7100} (\bibinfo {year} {2000})}\BibitemShut {NoStop}%
\bibitem [{\citenamefont {Nagato}\ and\ \citenamefont
	{Nagai}(1995)}]{Nagai_1995}%
\BibitemOpen
\bibfield  {author} {\bibinfo {author} {\bibfnamefont {Yasushi}\ \bibnamefont
		{Nagato}}\ and\ \bibinfo {author} {\bibfnamefont {Katsuhiko}\ \bibnamefont
		{Nagai}},\ }\bibfield  {title} {\enquote {\bibinfo {title} {Surface and size
			effect of a ${\mathit{d}}_{\mathit{x}\mathit{y}}$-state superconductor},}\
}\href {\doibase 10.1103/PhysRevB.51.16254} {\bibfield  {journal} {\bibinfo
	{journal} {Phys. Rev. B}\ }\textbf {\bibinfo {volume} {51}},\ \bibinfo
{pages} {16254--16258} (\bibinfo {year} {1995})}\BibitemShut {NoStop}%
\bibitem [{\citenamefont {Matsumoto}\ and\ \citenamefont
	{Shiba}(1996)}]{Shiba_1996}%
\BibitemOpen
\bibfield  {author} {\bibinfo {author} {\bibfnamefont {Masashige}\
		\bibnamefont {Matsumoto}}\ and\ \bibinfo {author} {\bibfnamefont {Hiroyuki}\
		\bibnamefont {Shiba}},\ }\bibfield  {title} {\enquote {\bibinfo {title}
		{Coexistence of different symmetry order parameters near a surface in d-wave
			superconductors iii},}\ }\href {\doibase 10.1143/JPSJ.65.2194} {\bibfield
	{journal} {\bibinfo  {journal} {Journal of the Physical Society of Japan}\
	}\textbf {\bibinfo {volume} {65}},\ \bibinfo {pages} {2194--2203} (\bibinfo
	{year} {1996})},\ \Eprint
{http://arxiv.org/abs/http://dx.doi.org/10.1143/JPSJ.65.2194}
{http://dx.doi.org/10.1143/JPSJ.65.2194} \BibitemShut {NoStop}%
\bibitem [{\citenamefont {Eschrig}(2000)}]{Eschrig_2000}%
\BibitemOpen
\bibfield  {author} {\bibinfo {author} {\bibfnamefont {Matthias}\
		\bibnamefont {Eschrig}},\ }\bibfield  {title} {\enquote {\bibinfo {title}
		{Distribution functions in nonequilibrium theory of superconductivity and
			andreev spectroscopy in unconventional superconductors},}\ }\href {\doibase
	10.1103/PhysRevB.61.9061} {\bibfield  {journal} {\bibinfo  {journal} {Phys.
			Rev. B}\ }\textbf {\bibinfo {volume} {61}},\ \bibinfo {pages} {9061--9076}
	(\bibinfo {year} {2000})}\BibitemShut {NoStop}%
\bibitem [{\citenamefont {Fogelstr\"om}(2000)}]{Fogelstrom_2000}%
\BibitemOpen
\bibfield  {author} {\bibinfo {author} {\bibfnamefont {Mikael}\ \bibnamefont
		{Fogelstr\"om}},\ }\bibfield  {title} {\enquote {\bibinfo {title} {Josephson
			currents through spin-active interfaces},}\ }\href {\doibase
	10.1103/PhysRevB.62.11812} {\bibfield  {journal} {\bibinfo  {journal} {Phys.
			Rev. B}\ }\textbf {\bibinfo {volume} {62}},\ \bibinfo {pages} {11812--11819}
	(\bibinfo {year} {2000})}\BibitemShut {NoStop}%
\bibitem [{\citenamefont {Zhao}\ \emph {et~al.}(2004)\citenamefont {Zhao},
	\citenamefont {L\"ofwander},\ and\ \citenamefont {Sauls}}]{Sauls_2004}%
\BibitemOpen
\bibfield  {author} {\bibinfo {author} {\bibfnamefont {Erhai}\ \bibnamefont
		{Zhao}}, \bibinfo {author} {\bibfnamefont {Tomas}\ \bibnamefont
		{L\"ofwander}}, \ and\ \bibinfo {author} {\bibfnamefont {J.~A.}\ \bibnamefont
		{Sauls}},\ }\bibfield  {title} {\enquote {\bibinfo {title} {Nonequilibrium
			superconductivity near spin-active interfaces},}\ }\href {\doibase
	10.1103/PhysRevB.70.134510} {\bibfield  {journal} {\bibinfo  {journal} {Phys.
			Rev. B}\ }\textbf {\bibinfo {volume} {70}},\ \bibinfo {pages} {134510}
	(\bibinfo {year} {2004})}\BibitemShut {NoStop}%
\bibitem [{\citenamefont {Blonder}\ \emph {et~al.}(1982)\citenamefont
	{Blonder}, \citenamefont {Tinkham},\ and\ \citenamefont {Klapwijk}}]{BTK}%
\BibitemOpen
\bibfield  {author} {\bibinfo {author} {\bibfnamefont {G.~E.}\ \bibnamefont
		{Blonder}}, \bibinfo {author} {\bibfnamefont {M.}~\bibnamefont {Tinkham}}, \
	and\ \bibinfo {author} {\bibfnamefont {T.~M.}\ \bibnamefont {Klapwijk}},\
}\bibfield  {title} {\enquote {\bibinfo {title} {Transition from metallic to
		tunneling regimes in superconducting microconstrictions: Excess current,
		charge imbalance, and supercurrent conversion},}\ }\href {\doibase
10.1103/PhysRevB.25.4515} {\bibfield  {journal} {\bibinfo  {journal} {Phys.
		Rev. B}\ }\textbf {\bibinfo {volume} {25}},\ \bibinfo {pages} {4515--4532}
(\bibinfo {year} {1982})}\BibitemShut {NoStop}%
\bibitem [{Note2()}]{Note2}%
\BibitemOpen
\bibinfo {note} {In the self-consistent evaluation of the pairing states, we
	have not included thermodynamic phenomena. Accordingly, in the numerical
	calculations, we choose a sufficiently small temperature
	$T=0.05T_c$.}\BibitemShut {Stop}%
\bibitem [{\citenamefont {Zhu}\ and\ \citenamefont
	{Ting}(1999)}]{Zhu-Ting_1999}%
\BibitemOpen
\bibfield  {author} {\bibinfo {author} {\bibfnamefont {Jian-Xin}\
		\bibnamefont {Zhu}}\ and\ \bibinfo {author} {\bibfnamefont {C.~S.}\
		\bibnamefont {Ting}},\ }\bibfield  {title} {\enquote {\bibinfo {title} {Shot
			noise in a normal-metal\char21{}$d$-wave superconductor junction with a
			${110}$-oriented interface},}\ }\href {\doibase 10.1103/PhysRevB.59.R14165}
{\bibfield  {journal} {\bibinfo  {journal} {Phys. Rev. B}\ }\textbf {\bibinfo
		{volume} {59}},\ \bibinfo {pages} {R14165--R14168} (\bibinfo {year}
	{1999})}\BibitemShut {NoStop}%
\bibitem [{\citenamefont {Tanaka}\ \emph {et~al.}(2000)\citenamefont {Tanaka},
	\citenamefont {Asai}, \citenamefont {Yoshida}, \citenamefont {Inoue},\ and\
	\citenamefont {Kashiwaya}}]{Inoue_2000}%
\BibitemOpen
\bibfield  {author} {\bibinfo {author} {\bibfnamefont {Y.}~\bibnamefont
		{Tanaka}}, \bibinfo {author} {\bibfnamefont {T.}~\bibnamefont {Asai}},
	\bibinfo {author} {\bibfnamefont {N.}~\bibnamefont {Yoshida}}, \bibinfo
	{author} {\bibfnamefont {J.}~\bibnamefont {Inoue}}, \ and\ \bibinfo {author}
	{\bibfnamefont {S.}~\bibnamefont {Kashiwaya}},\ }\bibfield  {title} {\enquote
	{\bibinfo {title} {Interface effects on the shot noise in
			normal-metal\char21{} \textit{d} -wave superconductor junctions},}\ }\href
{\doibase 10.1103/PhysRevB.61.R11902} {\bibfield  {journal} {\bibinfo
		{journal} {Phys. Rev. B}\ }\textbf {\bibinfo {volume} {61}},\ \bibinfo
	{pages} {R11902--R11905} (\bibinfo {year} {2000})}\BibitemShut {NoStop}%
\bibitem [{\citenamefont {Tanaka}\ and\ \citenamefont
	{Golubov}(2007)}]{Tanaka_2007}%
\BibitemOpen
\bibfield  {author} {\bibinfo {author} {\bibfnamefont {Y.}~\bibnamefont
		{Tanaka}}\ and\ \bibinfo {author} {\bibfnamefont {A.~A.}\ \bibnamefont
		{Golubov}},\ }\bibfield  {title} {\enquote {\bibinfo {title} {Theory of the
			proximity effect in junctions with unconventional superconductors},}\ }\href
{\doibase 10.1103/PhysRevLett.98.037003} {\bibfield  {journal} {\bibinfo
		{journal} {Phys. Rev. Lett.}\ }\textbf {\bibinfo {volume} {98}},\ \bibinfo
	{pages} {037003} (\bibinfo {year} {2007})}\BibitemShut {NoStop}%
\bibitem [{\citenamefont {Bergeret}\ \emph {et~al.}(2001)\citenamefont
	{Bergeret}, \citenamefont {Volkov},\ and\ \citenamefont
	{Efetov}}]{Bergeret_2001}%
\BibitemOpen
\bibfield  {author} {\bibinfo {author} {\bibfnamefont {F.~S.}\ \bibnamefont
		{Bergeret}}, \bibinfo {author} {\bibfnamefont {A.~F.}\ \bibnamefont
		{Volkov}}, \ and\ \bibinfo {author} {\bibfnamefont {K.~B.}\ \bibnamefont
		{Efetov}},\ }\bibfield  {title} {\enquote {\bibinfo {title} {Long-range
			proximity effects in superconductor-ferromagnet structures},}\ }\href
{\doibase 10.1103/PhysRevLett.86.4096} {\bibfield  {journal} {\bibinfo
		{journal} {Phys. Rev. Lett.}\ }\textbf {\bibinfo {volume} {86}},\ \bibinfo
	{pages} {4096--4099} (\bibinfo {year} {2001})}\BibitemShut {NoStop}%
\bibitem [{\citenamefont {Bergeret}\ \emph {et~al.}(2005)\citenamefont
	{Bergeret}, \citenamefont {Volkov},\ and\ \citenamefont
	{Efetov}}]{Bergeret_RMP}%
\BibitemOpen
\bibfield  {author} {\bibinfo {author} {\bibfnamefont {F.~S.}\ \bibnamefont
		{Bergeret}}, \bibinfo {author} {\bibfnamefont {A.~F.}\ \bibnamefont
		{Volkov}}, \ and\ \bibinfo {author} {\bibfnamefont {K.~B.}\ \bibnamefont
		{Efetov}},\ }\bibfield  {title} {\enquote {\bibinfo {title} {Odd triplet
			superconductivity and related phenomena in superconductor-ferromagnet
			structures},}\ }\href {\doibase 10.1103/RevModPhys.77.1321} {\bibfield
	{journal} {\bibinfo  {journal} {Rev. Mod. Phys.}\ }\textbf {\bibinfo {volume}
		{77}},\ \bibinfo {pages} {1321--1373} (\bibinfo {year} {2005})}\BibitemShut
{NoStop}%
\bibitem [{\citenamefont {Linder}\ \emph {et~al.}(2010)\citenamefont {Linder},
	\citenamefont {Sudb\o{}}, \citenamefont {Yokoyama}, \citenamefont {Grein},\
	and\ \citenamefont {Eschrig}}]{Linder_2010c}%
\BibitemOpen
\bibfield  {author} {\bibinfo {author} {\bibfnamefont {Jacob}\ \bibnamefont
		{Linder}}, \bibinfo {author} {\bibfnamefont {Asle}\ \bibnamefont {Sudb\o{}}},
	\bibinfo {author} {\bibfnamefont {Takehito}\ \bibnamefont {Yokoyama}},
	\bibinfo {author} {\bibfnamefont {Roland}\ \bibnamefont {Grein}}, \ and\
	\bibinfo {author} {\bibfnamefont {Matthias}\ \bibnamefont {Eschrig}},\
}\bibfield  {title} {\enquote {\bibinfo {title} {Signature of odd-frequency
		pairing correlations induced by a magnetic interface},}\ }\href {\doibase
10.1103/PhysRevB.81.214504} {\bibfield  {journal} {\bibinfo  {journal} {Phys.
		Rev. B}\ }\textbf {\bibinfo {volume} {81}},\ \bibinfo {pages} {214504}
(\bibinfo {year} {2010})}\BibitemShut {NoStop}%
\bibitem [{\citenamefont {Jacobsen}\ and\ \citenamefont
	{Linder}(2015)}]{Linder_2015}%
\BibitemOpen
\bibfield  {author} {\bibinfo {author} {\bibfnamefont {Sol~H.}\ \bibnamefont
		{Jacobsen}}\ and\ \bibinfo {author} {\bibfnamefont {Jacob}\ \bibnamefont
		{Linder}},\ }\bibfield  {title} {\enquote {\bibinfo {title} {Giant triplet
			proximity effect in $\ensuremath{\pi}$-biased josephson junctions with
			spin-orbit coupling},}\ }\href {\doibase 10.1103/PhysRevB.92.024501}
{\bibfield  {journal} {\bibinfo  {journal} {Phys. Rev. B}\ }\textbf {\bibinfo
		{volume} {92}},\ \bibinfo {pages} {024501} (\bibinfo {year}
	{2015})}\BibitemShut {NoStop}%
\bibitem [{\citenamefont {Gomperud}\ and\ \citenamefont
	{Linder}(2015)}]{Linder_2015b}%
\BibitemOpen
\bibfield  {author} {\bibinfo {author} {\bibfnamefont {Ingvild}\ \bibnamefont
		{Gomperud}}\ and\ \bibinfo {author} {\bibfnamefont {Jacob}\ \bibnamefont
		{Linder}},\ }\bibfield  {title} {\enquote {\bibinfo {title} {Spin
			supercurrent and phase-tunable triplet cooper pairs via magnetic
			insulators},}\ }\href {\doibase 10.1103/PhysRevB.92.035416} {\bibfield
	{journal} {\bibinfo  {journal} {Phys. Rev. B}\ }\textbf {\bibinfo {volume}
		{92}},\ \bibinfo {pages} {035416} (\bibinfo {year} {2015})}\BibitemShut
{NoStop}%
\bibitem [{\citenamefont {Tanaka}\ \emph
	{et~al.}(2007{\natexlab{b}})\citenamefont {Tanaka}, \citenamefont {Golubov},
	\citenamefont {Kashiwaya},\ and\ \citenamefont {Ueda}}]{Tanaka_2007b}%
\BibitemOpen
\bibfield  {author} {\bibinfo {author} {\bibfnamefont {Yukio}\ \bibnamefont
		{Tanaka}}, \bibinfo {author} {\bibfnamefont {Alexander~A.}\ \bibnamefont
		{Golubov}}, \bibinfo {author} {\bibfnamefont {Satoshi}\ \bibnamefont
		{Kashiwaya}}, \ and\ \bibinfo {author} {\bibfnamefont {Masahito}\
		\bibnamefont {Ueda}},\ }\bibfield  {title} {\enquote {\bibinfo {title}
		{Anomalous josephson effect between even- and odd-frequency
			superconductors},}\ }\href {\doibase 10.1103/PhysRevLett.99.037005}
{\bibfield  {journal} {\bibinfo  {journal} {Phys. Rev. Lett.}\ }\textbf
	{\bibinfo {volume} {99}},\ \bibinfo {pages} {037005} (\bibinfo {year}
	{2007}{\natexlab{b}})}\BibitemShut {NoStop}%
\bibitem [{\citenamefont {Yokoyama}(2012)}]{Yokoyama_2012}%
\BibitemOpen
\bibfield  {author} {\bibinfo {author} {\bibfnamefont {Takehito}\
		\bibnamefont {Yokoyama}},\ }\bibfield  {title} {\enquote {\bibinfo {title}
		{Josephson and proximity effects on the surface of a topological
			insulator},}\ }\href {\doibase 10.1103/PhysRevB.86.075410} {\bibfield
	{journal} {\bibinfo  {journal} {Phys. Rev. B}\ }\textbf {\bibinfo {volume}
		{86}},\ \bibinfo {pages} {075410} (\bibinfo {year} {2012})}\BibitemShut
{NoStop}%
\bibitem [{\citenamefont {Black-Schaffer}\ and\ \citenamefont
	{Balatsky}(2012)}]{Black-Schaffer_2012}%
\BibitemOpen
\bibfield  {author} {\bibinfo {author} {\bibfnamefont {Annica~M.}\
		\bibnamefont {Black-Schaffer}}\ and\ \bibinfo {author} {\bibfnamefont
		{Alexander~V.}\ \bibnamefont {Balatsky}},\ }\bibfield  {title} {\enquote
	{\bibinfo {title} {Odd-frequency superconducting pairing in topological
			insulators},}\ }\href {\doibase 10.1103/PhysRevB.86.144506} {\bibfield
	{journal} {\bibinfo  {journal} {Phys. Rev. B}\ }\textbf {\bibinfo {volume}
		{86}},\ \bibinfo {pages} {144506} (\bibinfo {year} {2012})}\BibitemShut
{NoStop}%
\bibitem [{\citenamefont {Cr\'epin}\ \emph {et~al.}(2015)\citenamefont
	{Cr\'epin}, \citenamefont {Burset},\ and\ \citenamefont
	{Trauzettel}}]{Crepin_2015}%
\BibitemOpen
\bibfield  {author} {\bibinfo {author} {\bibfnamefont {Fran\c{c}ois}\
		\bibnamefont {Cr\'epin}}, \bibinfo {author} {\bibfnamefont {Pablo}\
		\bibnamefont {Burset}}, \ and\ \bibinfo {author} {\bibfnamefont {Bj\"orn}\
		\bibnamefont {Trauzettel}},\ }\bibfield  {title} {\enquote {\bibinfo {title}
		{Odd-frequency triplet superconductivity at the helical edge of a topological
			insulator},}\ }\href {\doibase 10.1103/PhysRevB.92.100507} {\bibfield
	{journal} {\bibinfo  {journal} {Phys. Rev. B}\ }\textbf {\bibinfo {volume}
		{92}},\ \bibinfo {pages} {100507} (\bibinfo {year} {2015})}\BibitemShut
{NoStop}%
\bibitem [{\citenamefont {Burset}\ \emph {et~al.}(2015)\citenamefont {Burset},
	\citenamefont {Lu}, \citenamefont {Tkachov}, \citenamefont {Tanaka},
	\citenamefont {Hankiewicz},\ and\ \citenamefont {Trauzettel}}]{Burset_2015}%
\BibitemOpen
\bibfield  {author} {\bibinfo {author} {\bibfnamefont {Pablo}\ \bibnamefont
		{Burset}}, \bibinfo {author} {\bibfnamefont {Bo}~\bibnamefont {Lu}}, \bibinfo
	{author} {\bibfnamefont {Grigory}\ \bibnamefont {Tkachov}}, \bibinfo {author}
	{\bibfnamefont {Yukio}\ \bibnamefont {Tanaka}}, \bibinfo {author}
	{\bibfnamefont {Ewelina~M.}\ \bibnamefont {Hankiewicz}}, \ and\ \bibinfo
	{author} {\bibfnamefont {Bj\"orn}\ \bibnamefont {Trauzettel}},\ }\bibfield
{title} {\enquote {\bibinfo {title} {Superconducting proximity effect in
			three-dimensional topological insulators in the presence of a magnetic
			field},}\ }\href {\doibase 10.1103/PhysRevB.92.205424} {\bibfield  {journal}
	{\bibinfo  {journal} {Phys. Rev. B}\ }\textbf {\bibinfo {volume} {92}},\
	\bibinfo {pages} {205424} (\bibinfo {year} {2015})}\BibitemShut {NoStop}%
\bibitem [{\citenamefont {Aperis}\ \emph {et~al.}(2015)\citenamefont {Aperis},
	\citenamefont {Maldonado},\ and\ \citenamefont {Oppeneer}}]{Aperis_2015}%
\BibitemOpen
\bibfield  {author} {\bibinfo {author} {\bibfnamefont {Alex}\ \bibnamefont
		{Aperis}}, \bibinfo {author} {\bibfnamefont {Pablo}\ \bibnamefont
		{Maldonado}}, \ and\ \bibinfo {author} {\bibfnamefont {Peter~M.}\
		\bibnamefont {Oppeneer}},\ }\bibfield  {title} {\enquote {\bibinfo {title}
		{\textit{Ab initio} theory of magnetic-field-induced odd-frequency two-band
			superconductivity in ${\mathrm{mgb}}_{2}$},}\ }\href {\doibase
	10.1103/PhysRevB.92.054516} {\bibfield  {journal} {\bibinfo  {journal} {Phys.
			Rev. B}\ }\textbf {\bibinfo {volume} {92}},\ \bibinfo {pages} {054516}
	(\bibinfo {year} {2015})}\BibitemShut {NoStop}%
\bibitem [{\citenamefont {Sothmann}\ \emph {et~al.}(2014)\citenamefont
	{Sothmann}, \citenamefont {Weiss}, \citenamefont {Governale},\ and\
	\citenamefont {K\"onig}}]{Sothmann_2014}%
\BibitemOpen
\bibfield  {author} {\bibinfo {author} {\bibfnamefont {Bj\"orn}\ \bibnamefont
		{Sothmann}}, \bibinfo {author} {\bibfnamefont {Stephan}\ \bibnamefont
		{Weiss}}, \bibinfo {author} {\bibfnamefont {Michele}\ \bibnamefont
		{Governale}}, \ and\ \bibinfo {author} {\bibfnamefont {J\"urgen}\
		\bibnamefont {K\"onig}},\ }\bibfield  {title} {\enquote {\bibinfo {title}
		{Unconventional superconductivity in double quantum dots},}\ }\href {\doibase
	10.1103/PhysRevB.90.220501} {\bibfield  {journal} {\bibinfo  {journal} {Phys.
			Rev. B}\ }\textbf {\bibinfo {volume} {90}},\ \bibinfo {pages} {220501}
	(\bibinfo {year} {2014})}\BibitemShut {NoStop}%
\bibitem [{\citenamefont {Burset}\ \emph {et~al.}(2016)\citenamefont {Burset},
	\citenamefont {Lu}, \citenamefont {Ebisu}, \citenamefont {Asano},\ and\
	\citenamefont {Tanaka}}]{Burset_2016}%
\BibitemOpen
\bibfield  {author} {\bibinfo {author} {\bibfnamefont {Pablo}\ \bibnamefont
		{Burset}}, \bibinfo {author} {\bibfnamefont {Bo}~\bibnamefont {Lu}}, \bibinfo
	{author} {\bibfnamefont {Hiromi}\ \bibnamefont {Ebisu}}, \bibinfo {author}
	{\bibfnamefont {Yasuhiro}\ \bibnamefont {Asano}}, \ and\ \bibinfo {author}
	{\bibfnamefont {Yukio}\ \bibnamefont {Tanaka}},\ }\bibfield  {title}
{\enquote {\bibinfo {title} {All-electrical generation and control of
			odd-frequency $s$-wave cooper pairs in double quantum dots},}\ }\href
{\doibase 10.1103/PhysRevB.93.201402} {\bibfield  {journal} {\bibinfo
		{journal} {Phys. Rev. B}\ }\textbf {\bibinfo {volume} {93}},\ \bibinfo
	{pages} {201402} (\bibinfo {year} {2016})}\BibitemShut {NoStop}%
\bibitem [{\citenamefont {Black-Schaffer}\ and\ \citenamefont
	{Balatsky}(2013)}]{Black-Schaffer_2013b}%
\BibitemOpen
\bibfield  {author} {\bibinfo {author} {\bibfnamefont {Annica~M.}\
		\bibnamefont {Black-Schaffer}}\ and\ \bibinfo {author} {\bibfnamefont
		{Alexander~V.}\ \bibnamefont {Balatsky}},\ }\bibfield  {title} {\enquote
	{\bibinfo {title} {Odd-frequency superconducting pairing in multiband
			superconductors},}\ }\href {\doibase 10.1103/PhysRevB.88.104514} {\bibfield
	{journal} {\bibinfo  {journal} {Phys. Rev. B}\ }\textbf {\bibinfo {volume}
		{88}},\ \bibinfo {pages} {104514} (\bibinfo {year} {2013})}\BibitemShut
{NoStop}%
\bibitem [{\citenamefont {Asano}\ and\ \citenamefont
	{Sasaki}(2015)}]{Asano_2015}%
\BibitemOpen
\bibfield  {author} {\bibinfo {author} {\bibfnamefont {Yasuhiro}\
		\bibnamefont {Asano}}\ and\ \bibinfo {author} {\bibfnamefont {Akihiro}\
		\bibnamefont {Sasaki}},\ }\bibfield  {title} {\enquote {\bibinfo {title}
		{Odd-frequency cooper pairs in two-band superconductors and their magnetic
			response},}\ }\href {\doibase 10.1103/PhysRevB.92.224508} {\bibfield
	{journal} {\bibinfo  {journal} {Phys. Rev. B}\ }\textbf {\bibinfo {volume}
		{92}},\ \bibinfo {pages} {224508} (\bibinfo {year} {2015})}\BibitemShut
{NoStop}%
\bibitem [{\citenamefont {Tanaka}\ \emph {et~al.}(2012)\citenamefont {Tanaka},
	\citenamefont {Sato},\ and\ \citenamefont {Nagaosa}}]{Tanaka_JPSJ}%
\BibitemOpen
\bibfield  {author} {\bibinfo {author} {\bibfnamefont {Yukio}\ \bibnamefont
		{Tanaka}}, \bibinfo {author} {\bibfnamefont {Masatoshi}\ \bibnamefont
		{Sato}}, \ and\ \bibinfo {author} {\bibfnamefont {Naoto}\ \bibnamefont
		{Nagaosa}},\ }\bibfield  {title} {\enquote {\bibinfo {title} {Symmetry and
			topology in superconductors ---odd-frequency pairing and edge states---},}\
}\href {\doibase 10.1143/JPSJ.81.011013} {\bibfield  {journal} {\bibinfo
	{journal} {Journal of the Physical Society of Japan}\ }\textbf {\bibinfo
	{volume} {81}},\ \bibinfo {pages} {011013} (\bibinfo {year}
{2012})}\BibitemShut {NoStop}%
\bibitem [{\citenamefont {Eschrig}(2015)}]{Eschrig_RPP}%
\BibitemOpen
\bibfield  {author} {\bibinfo {author} {\bibfnamefont {Matthias}\
		\bibnamefont {Eschrig}},\ }\bibfield  {title} {\enquote {\bibinfo {title}
		{Spin-polarized supercurrents for spintronics: a review of current
			progress},}\ }\href {http://stacks.iop.org/0034-4885/78/i=10/a=104501}
{\bibfield  {journal} {\bibinfo  {journal} {Reports on Progress in Physics}\
	}\textbf {\bibinfo {volume} {78}},\ \bibinfo {pages} {104501} (\bibinfo
	{year} {2015})}\BibitemShut {NoStop}%
\bibitem [{\citenamefont {Gnezdilov}\ \emph {et~al.}(2015)\citenamefont
	{Gnezdilov}, \citenamefont {van Heck}, \citenamefont {Diez}, \citenamefont
	{Hutasoit},\ and\ \citenamefont {Beenakker}}]{Beenakker_2015}%
\BibitemOpen
\bibfield  {author} {\bibinfo {author} {\bibfnamefont {N.~V.}\ \bibnamefont
		{Gnezdilov}}, \bibinfo {author} {\bibfnamefont {B.}~\bibnamefont {van Heck}},
	\bibinfo {author} {\bibfnamefont {M.}~\bibnamefont {Diez}}, \bibinfo {author}
	{\bibfnamefont {Jimmy~A.}\ \bibnamefont {Hutasoit}}, \ and\ \bibinfo {author}
	{\bibfnamefont {C.~W.~J.}\ \bibnamefont {Beenakker}},\ }\bibfield  {title}
{\enquote {\bibinfo {title} {Topologically protected charge transfer along
			the edge of a chiral $p$-wave superconductor},}\ }\href {\doibase
	10.1103/PhysRevB.92.121406} {\bibfield  {journal} {\bibinfo  {journal} {Phys.
			Rev. B}\ }\textbf {\bibinfo {volume} {92}},\ \bibinfo {pages} {121406}
	(\bibinfo {year} {2015})}\BibitemShut {NoStop}%
\bibitem [{\citenamefont {Cuevas}\ \emph {et~al.}(1996)\citenamefont {Cuevas},
	\citenamefont {Mart\'{\i}n-Rodero},\ and\ \citenamefont
	{Yeyati}}]{Cuevas_1996}%
\BibitemOpen
\bibfield  {author} {\bibinfo {author} {\bibfnamefont {J.~C.}\ \bibnamefont
		{Cuevas}}, \bibinfo {author} {\bibfnamefont {A.}~\bibnamefont
		{Mart\'{\i}n-Rodero}}, \ and\ \bibinfo {author} {\bibfnamefont {A.~Levy}\
		\bibnamefont {Yeyati}},\ }\bibfield  {title} {\enquote {\bibinfo {title}
		{Hamiltonian approach to the transport properties of superconducting quantum
			point contacts},}\ }\href {\doibase 10.1103/PhysRevB.54.7366} {\bibfield
	{journal} {\bibinfo  {journal} {Phys. Rev. B}\ }\textbf {\bibinfo {volume}
		{54}},\ \bibinfo {pages} {7366--7379} (\bibinfo {year} {1996})}\BibitemShut
{NoStop}%
\bibitem [{\citenamefont {Bakurskiy}\ \emph {et~al.}(2014)\citenamefont
	{Bakurskiy}, \citenamefont {Golubov}, \citenamefont {Kupriyanov},
	\citenamefont {Yada},\ and\ \citenamefont {Tanaka}}]{Bakurskiy_2014}%
\BibitemOpen
\bibfield  {author} {\bibinfo {author} {\bibfnamefont {S.~V.}\ \bibnamefont
		{Bakurskiy}}, \bibinfo {author} {\bibfnamefont {A.~A.}\ \bibnamefont
		{Golubov}}, \bibinfo {author} {\bibfnamefont {M.~Yu.}\ \bibnamefont
		{Kupriyanov}}, \bibinfo {author} {\bibfnamefont {K.}~\bibnamefont {Yada}}, \
	and\ \bibinfo {author} {\bibfnamefont {Y.}~\bibnamefont {Tanaka}},\
}\bibfield  {title} {\enquote {\bibinfo {title} {Anomalous surface states at
		interfaces in $p$-wave superconductors},}\ }\href {\doibase
10.1103/PhysRevB.90.064513} {\bibfield  {journal} {\bibinfo  {journal} {Phys.
		Rev. B}\ }\textbf {\bibinfo {volume} {90}},\ \bibinfo {pages} {064513}
(\bibinfo {year} {2014})}\BibitemShut {NoStop}%
\bibitem [{Note3()}]{Note3}%
\BibitemOpen
\bibinfo {note} {The noise-current ratio in Fig.~\ref {fig:ivc}(d-f) is only
	well defined for finite voltage, since $I_S(V\protect \tmspace -\thinmuskip
	{.1667em}=\protect \tmspace -\thinmuskip {.1667em}0)\protect \tmspace
	-\thinmuskip {.1667em}=\protect \tmspace -\thinmuskip
	{.1667em}0$.}\BibitemShut {Stop}%
\bibitem [{\citenamefont {Scheer}\ \emph {et~al.}(2001)\citenamefont {Scheer},
	\citenamefont {Belzig}, \citenamefont {Naveh}, \citenamefont {Devoret},
	\citenamefont {Esteve},\ and\ \citenamefont {Urbina}}]{Scheer_2001}%
\BibitemOpen
\bibfield  {author} {\bibinfo {author} {\bibfnamefont {E.}~\bibnamefont
		{Scheer}}, \bibinfo {author} {\bibfnamefont {W.}~\bibnamefont {Belzig}},
	\bibinfo {author} {\bibfnamefont {Y.}~\bibnamefont {Naveh}}, \bibinfo
	{author} {\bibfnamefont {M.~H.}\ \bibnamefont {Devoret}}, \bibinfo {author}
	{\bibfnamefont {D.}~\bibnamefont {Esteve}}, \ and\ \bibinfo {author}
	{\bibfnamefont {C.}~\bibnamefont {Urbina}},\ }\bibfield  {title} {\enquote
	{\bibinfo {title} {Proximity effect and multiple andreev reflections in gold
			atomic contacts},}\ }\href {\doibase 10.1103/PhysRevLett.86.284} {\bibfield
	{journal} {\bibinfo  {journal} {Phys. Rev. Lett.}\ }\textbf {\bibinfo
		{volume} {86}},\ \bibinfo {pages} {284--287} (\bibinfo {year}
	{2001})}\BibitemShut {NoStop}%
\bibitem [{\citenamefont {Wiedenmann}\ \emph {et~al.}(2016)\citenamefont
	{Wiedenmann}, \citenamefont {Bocquillon}, \citenamefont {Deacon},
	\citenamefont {Hartinger}, \citenamefont {Herrmann}, \citenamefont
	{Klapwijk}, \citenamefont {Maier}, \citenamefont {Ames}, \citenamefont
	{Brüne}, \citenamefont {Gould}, \citenamefont {Oiwa}, \citenamefont
	{Ishibashi}, \citenamefont {Tarucha}, \citenamefont {Buhmann},\ and\
	\citenamefont {Molenkamp}}]{Jonas_2016}%
\BibitemOpen
\bibfield  {author} {\bibinfo {author} {\bibfnamefont {J.}~\bibnamefont
		{Wiedenmann}}, \bibinfo {author} {\bibfnamefont {E.}~\bibnamefont
		{Bocquillon}}, \bibinfo {author} {\bibfnamefont {R.~S.}\ \bibnamefont
		{Deacon}}, \bibinfo {author} {\bibfnamefont {S.}~\bibnamefont {Hartinger}},
	\bibinfo {author} {\bibfnamefont {O.}~\bibnamefont {Herrmann}}, \bibinfo
	{author} {\bibfnamefont {T.~M.}\ \bibnamefont {Klapwijk}}, \bibinfo {author}
	{\bibfnamefont {L.}~\bibnamefont {Maier}}, \bibinfo {author} {\bibfnamefont
		{C.}~\bibnamefont {Ames}}, \bibinfo {author} {\bibfnamefont {C.}~\bibnamefont
		{Brüne}}, \bibinfo {author} {\bibfnamefont {C.}~\bibnamefont {Gould}},
	\bibinfo {author} {\bibfnamefont {A.}~\bibnamefont {Oiwa}}, \bibinfo {author}
	{\bibfnamefont {K.}~\bibnamefont {Ishibashi}}, \bibinfo {author}
	{\bibfnamefont {S.}~\bibnamefont {Tarucha}}, \bibinfo {author} {\bibfnamefont
		{H.}~\bibnamefont {Buhmann}}, \ and\ \bibinfo {author} {\bibfnamefont
		{L.~W.}\ \bibnamefont {Molenkamp}},\ }\bibfield  {title} {\enquote {\bibinfo
		{title} {4$\ensuremath{\pi}$-periodic josephson supercurrent in hgte-based
			topological josephson junctions},}\ }\href {\doibase 10.1038/ncomms10303}
{\bibfield  {journal} {\bibinfo  {journal} {Nature Communications}\ }\textbf
	{\bibinfo {volume} {7}},\ \bibinfo {pages} {10303} (\bibinfo {year}
	{2016})}\BibitemShut {NoStop}%
\bibitem [{\citenamefont {Tikhonov}\ \emph {et~al.}(2016)\citenamefont
	{Tikhonov}, \citenamefont {Shovkun}, \citenamefont {Snelder}, \citenamefont
	{Stehno}, \citenamefont {Huang}, \citenamefont {Golden}, \citenamefont
	{Golubov}, \citenamefont {Brinkman},\ and\ \citenamefont
	{Khrapai}}]{Brinkman_2016}%
\BibitemOpen
\bibfield  {author} {\bibinfo {author} {\bibfnamefont {E.~S.}\ \bibnamefont
		{Tikhonov}}, \bibinfo {author} {\bibfnamefont {D.~V.}\ \bibnamefont
		{Shovkun}}, \bibinfo {author} {\bibfnamefont {M.}~\bibnamefont {Snelder}},
	\bibinfo {author} {\bibfnamefont {M.~P.}\ \bibnamefont {Stehno}}, \bibinfo
	{author} {\bibfnamefont {Y.}~\bibnamefont {Huang}}, \bibinfo {author}
	{\bibfnamefont {M.~S.}\ \bibnamefont {Golden}}, \bibinfo {author}
	{\bibfnamefont {A.~A.}\ \bibnamefont {Golubov}}, \bibinfo {author}
	{\bibfnamefont {A.}~\bibnamefont {Brinkman}}, \ and\ \bibinfo {author}
	{\bibfnamefont {V.~S.}\ \bibnamefont {Khrapai}},\ }\bibfield  {title}
{\enquote {\bibinfo {title} {Andreev reflection in an $s$-type superconductor
			proximized 3d topological insulator},}\ }\href {\doibase
	10.1103/PhysRevLett.117.147001} {\bibfield  {journal} {\bibinfo  {journal}
		{Phys. Rev. Lett.}\ }\textbf {\bibinfo {volume} {117}},\ \bibinfo {pages}
	{147001} (\bibinfo {year} {2016})}\BibitemShut {NoStop}%
\bibitem [{\citenamefont {Tanaka}\ and\ \citenamefont
	{Kashiwaya}(2004)}]{Kashiwaya_2004}%
\BibitemOpen
\bibfield  {author} {\bibinfo {author} {\bibfnamefont {Y.}~\bibnamefont
		{Tanaka}}\ and\ \bibinfo {author} {\bibfnamefont {S.}~\bibnamefont
		{Kashiwaya}},\ }\bibfield  {title} {\enquote {\bibinfo {title} {Anomalous
			charge transport in triplet superconductor junctions},}\ }\href {\doibase
	10.1103/PhysRevB.70.012507} {\bibfield  {journal} {\bibinfo  {journal} {Phys.
			Rev. B}\ }\textbf {\bibinfo {volume} {70}},\ \bibinfo {pages} {012507}
	(\bibinfo {year} {2004})}\BibitemShut {NoStop}%
\bibitem [{\citenamefont {Tanaka}\ \emph {et~al.}(2005)\citenamefont {Tanaka},
	\citenamefont {Kashiwaya},\ and\ \citenamefont {Yokoyama}}]{Yokoyama_2005}%
\BibitemOpen
\bibfield  {author} {\bibinfo {author} {\bibfnamefont {Y.}~\bibnamefont
		{Tanaka}}, \bibinfo {author} {\bibfnamefont {S.}~\bibnamefont {Kashiwaya}}, \
	and\ \bibinfo {author} {\bibfnamefont {T.}~\bibnamefont {Yokoyama}},\
}\bibfield  {title} {\enquote {\bibinfo {title} {Theory of enhanced proximity
		effect by midgap andreev resonant state in diffusive normal-metal/triplet
		superconductor junctions},}\ }\href {\doibase 10.1103/PhysRevB.71.094513}
{\bibfield  {journal} {\bibinfo  {journal} {Phys. Rev. B}\ }\textbf {\bibinfo
		{volume} {71}},\ \bibinfo {pages} {094513} (\bibinfo {year}
	{2005})}\BibitemShut {NoStop}%
\bibitem [{\citenamefont {Ikegaya}\ \emph {et~al.}(2016)\citenamefont
	{Ikegaya}, \citenamefont {Suzuki}, \citenamefont {Tanaka},\ and\
	\citenamefont {Asano}}]{Ikegaya_2016}%
\BibitemOpen
\bibfield  {author} {\bibinfo {author} {\bibfnamefont {Satoshi}\ \bibnamefont
		{Ikegaya}}, \bibinfo {author} {\bibfnamefont {Shu-Ichiro}\ \bibnamefont
		{Suzuki}}, \bibinfo {author} {\bibfnamefont {Yukio}\ \bibnamefont {Tanaka}},
	\ and\ \bibinfo {author} {\bibfnamefont {Yasuhiro}\ \bibnamefont {Asano}},\
}\bibfield  {title} {\enquote {\bibinfo {title} {Quantization of conductance
		minimum and index theorem},}\ }\href {\doibase 10.1103/PhysRevB.94.054512}
{\bibfield  {journal} {\bibinfo  {journal} {Phys. Rev. B}\ }\textbf {\bibinfo
		{volume} {94}},\ \bibinfo {pages} {054512} (\bibinfo {year}
	{2016})}\BibitemShut {NoStop}%
\bibitem [{\citenamefont {Asano}\ and\ \citenamefont
	{Tanaka}(2013)}]{Asano_2013}%
\BibitemOpen
\bibfield  {author} {\bibinfo {author} {\bibfnamefont {Yasuhiro}\
		\bibnamefont {Asano}}\ and\ \bibinfo {author} {\bibfnamefont {Yukio}\
		\bibnamefont {Tanaka}},\ }\bibfield  {title} {\enquote {\bibinfo {title}
		{Majorana fermions and odd-frequency cooper pairs in a normal-metal nanowire
			proximity-coupled to a topological superconductor},}\ }\href {\doibase
	10.1103/PhysRevB.87.104513} {\bibfield  {journal} {\bibinfo  {journal} {Phys.
			Rev. B}\ }\textbf {\bibinfo {volume} {87}},\ \bibinfo {pages} {104513}
	(\bibinfo {year} {2013})}\BibitemShut {NoStop}%
\bibitem [{\citenamefont {Kashuba}\ \emph {et~al.}(2017)\citenamefont
	{Kashuba}, \citenamefont {Sothmann}, \citenamefont {Burset},\ and\
	\citenamefont {Trauzettel}}]{Kashuba_2017}%
\BibitemOpen
\bibfield  {author} {\bibinfo {author} {\bibfnamefont {O.}~\bibnamefont
		{Kashuba}}, \bibinfo {author} {\bibfnamefont {B.}~\bibnamefont {Sothmann}},
	\bibinfo {author} {\bibfnamefont {P.}~\bibnamefont {Burset}}, \ and\ \bibinfo
	{author} {\bibfnamefont {B.}~\bibnamefont {Trauzettel}},\ }\href
{http://arxiv.org/abs/1612.03356} {\enquote {\bibinfo {title} {The majorana
			stm as a perfect detector of odd-frequency superconductivity},}\ } (\bibinfo
{year} {2017}),\ \bibinfo {note} {arXiv:1612.03356}\BibitemShut {NoStop}%
\end{thebibliography}
\end{document}